\documentclass[]{aa}
\usepackage{graphicx}
\usepackage{natbib}
\bibpunct{(}{)}{;}{a}{}{,} 
\usepackage{txfonts}
\newcommand{\ghz}[1]          {$#1\:\mathrm{\:GHz}$}

\newcommand{\degg}[1]         {${#1}^{\circ}$}

\newcommand{\kms}[1]          {$#1\:\mathrm{\:km\, s^{-1}}$}

\newcommand{\mj}              {$\mathrm{m_{\mathrm{J}}}$}
\newcommand{\mh}              {$\mathrm{m_{\mathrm{H}}}$}
\newcommand{\mk}              {$\mathrm{m_{\mathrm{K}}}$}

\newcommand{\sfreq}[1]        {$#1\ \mathrm{\:arcsec^{-1}}$}
\newcommand{\mum}[1]          {$#1\ \mathrm{\mu m}$}

\newcommand{\anyapp}[2]       {$\sim #1$~#2}
\newcommand{\pa}[1]           {PA\,$\sim {#1}^{\circ}$}
\newcommand{\angpm}[2]        {${#1}^{\circ} \pm {#2}^{\circ}$}
\newcommand{\anypm}[2]        {$#1 \pm #2$}

\newcommand{\fovs}[2]         {$#1 \times #2$}
\newcommand{\fovp}[2]         {$#1\times#2$}

\newcommand{\lexpsun}[2]      {$#1 \times 10^{#2}\ \mathrm{L}_{\odot}$}
\newcommand{\lsuno}           {$\mathrm{L}_{\odot}$}

\newcommand{\avio}[1]         {$\mathrm{A}_{\mathrm{v}}$}
\newcommand{\akapp}[1]        {$\mathrm{A}_{\mathrm{K}}$}
\newcommand{\kmsc}            {$\mathrm{kms}^{-1}$}

\newcommand{\hii}             {H{\sc ii}}

\newcommand{\hal}             {$\mathrm{H\alpha}$}

\newcommand{\brq}             {$\mathrm{Br}\gamma$}

\newcommand{\molh}            {$\mathrm{H}_{2}$}  
\newcommand{\molhl}           {$\mathrm{H_{2}\:(1 - 0)\:S(1)}$}

\newcommand{\thco}            {$\mathrm{^{13}CO}$} 
\newcommand{\cseveno}         {$\mathrm{C^{17}O}$}

\newcommand{\eng}[2]          {$#1\times10^{#2}$}

\newcommand{\texpo}           {$\mathrm{t_{exp}}$}

\newcommand{\fluunit}[2]      {$\mathrm{erg\ s}^{-1} \mathrm{cm}^{-2}$}

\newcommand{\ras}[1]          {$\mathrm{\alpha\ (h m s)}$}
\newcommand{\dec}[1]          {$\mathrm{\delta\ (^{\circ} {'} {''})}$}

\newcommand{\glftn}           {\object{GL\,4029\,IRS1}}

\newcommand{\lkhao}           {\object{LkH$\alpha$\,101}}

\newcommand{\glf}             {\object{GL\,490}}

\newcommand{\gla}             {\object{GL\,961}}
\newcommand{\glae}            {\object{GL\,961-E}}
\newcommand{\glaw}            {\object{GL\,961-W}}
\newcommand{\glawa}           {\object{GL\,961-Wa}}
\newcommand{\glawb}           {\object{GL\,961-Wb}}
\newcommand{\glfts}           {\object{GL\,437}}
\newcommand{\glftsn}          {\object{GL\,437-N}}

\newcommand{\glftss}          {\object{GL\,437-S}}

\newcommand{\fan}             {\object{GL\,961-Fan}}
\newcommand{\glb}             {\object{GL\,989}}

\newcommand{\glt}             {\object{GL\,2591}}

\newcommand{\mon}             {\object{Mon\,R2\,IRS3}}

\newcommand{\mont}            {\object{Mon\,R2\,IRS2}}

\newcommand{\monr}            {\object{Mon\,R2}}
\newcommand{\stwot}           {\object{S255\,IRS3}}

\newcommand{\stwo}            {\object{S255\,IRS1}}

\newcommand{\sonef}           {\object{S140\,IRS1}}

\newcommand{\sir}             {\object{S106\,IR}}

\newcommand{\ngcsfte}         {\object{NGC\,7538\,IRS1}}

\newcommand{\sone}            {\object{S235\,IRS1}}

\newcommand{\sttft}           {\object{S235\,IRS2}}

\newcommand{\sfour}           {\object{S235\,IRS4}}

\newcommand{\ngctwo}          {\object{NGC\,2024\,IRS2}}

\newcommand{\gfive}           {\object{GL\,5180\,IRS1}}

\begin{document}
   \title{Near-IR speckle imaging of massive young stellar objects}

   \subtitle{}

   \author{C. Alvarez
          \inst{1,2,3}
          \and
          M. Hoare\inst{1}
          \and
          A. Glindemann\inst{4}
          \and
          A. Richichi\inst{4}
          }

   \offprints{C. Alvarez, \email{alvarez@mpia-hd.mpg.de}}

   \institute{School of Physics and Astronomy, University of Leeds,
         Leeds LS2 9JT, United Kingdom
         \and
         Kapteyn Astronomical Institute, Postbus 800, 9700 AV Groningen, 
         The Netherlands
         \and
	 Max-Planck-Institut f\"ur Astronomie, K\"onigstuhl 17,  
	 D-69117 Heidelberg,  Germany
	 \and
         European Southern Observatory, Karl-Schwarzschildstr. 2, 
	 85748 Garching bei M\"unchen, Germany 
   }

   \date{Received September 15, 1996; accepted March 16, 1997}

   \abstract{We present near-IR speckle images of 21 massive
     Young Stellar Objects (YSOs) associated with outflows. The aim of
     this study is to search for sub-arcsecond reflection nebulae 
     associated with the outflow cavity. We find that 6 of the massive 
     YSOs show a conical nebula which can be interpreted in
     terms of reflected light from the dusty walls of the outflow
     cavity. In all cases, the small scale structures seen in our
     images are compared with outflow indicators found in the
     literature. No clear correlation is found between the presence of the
     reflection nebulosity and any property such as degree of embeddedness.
     We also note that 3 of the sources show close companions, one of
     them belonging also to the sample with conical nebula. 
   }

   \maketitle
%

\section{Introduction}
\label{IntroductionSection}

Outflow is invariably associated with star formation and is likely
to play a key role in the process. It can take the form of
highly collimated jets that may help solve the angular momentum
problem \citep{Bacciotti02}. A wide-angle wind
could also be present \citep{Lee01} and these
phenomena then drive the bipolar molecular outflows in the ambient
cloud. Magnetohydrodynamic mechanisms have been proposed to explain
the driving and collimation of these outflows
(e.g. \citealp{GalliShu93,Tomisaka02}). This picture has mostly been
developed for low-mass star formation. 

There are good reasons for supposing that this may not apply to the
formation of the more massive OB stars. Although bipolar molecular
flows are just as common, there are very few cases where highly
collimated jets are seen. On the contrary, high resolution radio
imaging has revealed a few cases where ionised winds are equatorial
in nature, i.e. perpendicular to the bipolar molecular outflow
\citep{Hoare94,HoareMuxlow96,Hoare02}. 

From a theoretical point of view, we may also expect that radiation
pressure from these luminous, hot stars may play a larger role than
magnetic pressure. Indeed, radiation pressure can naturally account for
an equatorial wind being driven off the surface of an accretion disc
\citep{ProgaDrew98}. Hydrodynamic collimation could provide an
  alternative to magnetic models for the larger scale bipolar
  molecular flows \citep[e.g. ][]{YorkeWelz96,MellemaFrank97}. 

One approach to achieve a better insight into these mechanisms is to
study the circumstellar matter distribution at the scales where the 
outflow is likely to be driven and collimated. In particular, the 
morphology of the cavity opened by the outflow at scales of 
a few 100~AU may yield hints about the outflow driving mechanism. 
For instance, a wider cavity would be expected if the outflow 
is rather equatorial at the base, while an initially jet-like 
outflow would produce a narrower cavity.

A better knowledge of the circumstellar density distribution
morphology can be inferred from the near-IR observations of the
sub-arcsecond reflection nebula produced by light from the young star
that is scattered off the dust in the cavity walls. These nebulae have
already been observed in a few massive YSOs (\monr,
\citealp{Koresko93,Preibisch02}; \sonef, \citealp{Hoare96,Schertl00}). 
However, no study has covered a large sample of
sources. Observations of the sub-arcsecond morphology over a large
sample allows the questions of how frequent reflection
nebulae and close companions are amongst the massive YSOs to be
addressed. The
observed reflection nebulae can be compared with predictions from
radiative transfer models of the scattered light to constrain some of
the properties of the circumstellar density distribution, e.g. optical
depth, cavity shape and inclination angle 
\citep{WhitneyHartmann92,FischerHenning95,Lucas98,Wolf03,Whitney03,Alvarez04}.
 They can also be compared
to observations of the ionised wind morphology from radio continuum
observations at similarly high resolution.

Here, we present near-IR multi-wavelength speckle observations of the 
sub-arcsecond morphology on a sample of 21 massive YSOs. The targets 
are known to be associated with large scale bipolar outflows from
  molecular line and shocked molecular hydrogen 
  studies. They are nearby and bright enough for the speckle
method to be used. Some preliminary results of this work were shown in  
\citet{Hoare96}. In section~\ref{ObservationsSection} the observations 
and data reduction are described. The results for individual sources
are presented in Section~\ref{ResultsSection}. In
Section~\ref{DiscussionSection} the statistical properties of the sample 
(frequency of reflection nebulae, frequency of binaries) are 
discussed. Finally, some concluding remarks are presented in
Section~\ref{ConclusionsSection}.
   

\section[Observations]{Observations}
\label{ObservationsSection}
 
\subsection{UKIRT Observations}

These observations (see Table~\ref{TargetListTable}) were 
performed on December 16, and~17, 1997 with IRCAM3 at UKIRT. 
The \fovp{256}{256}\ InSb array was used with the $\times 5$\ 
magnifier giving a pixel scale of 0\farcs057~$\mathrm{pix}^{-1}$. 
Due to the fast readout required, for most of the sources, only the central 
\fovp{128}{128}\ pixels where used, covering a field of view of 
\fovs{7\farcs3}{7\farcs3}. The seeing conditions in the near-IR varied between 
$0\farcs6$\ and $1\farcs1$\ during the run. A minimum exposure time 
of 0.02~sec was used for the brighter objects, and up to 0.25~sec was 
used for the fainter ones, which is not short enough to completely freeze
the seeing, but still gives some high resolution information. Observations 
of the target were alternated, normally every 500 speckle snapshots, with 
similar 500~snapshots of a nearby reference star (within \degg{1} of the 
science object) and nearby sky. More than one reference star was
used where possible to allow for the possibility of undiscovered
multiplicity in them.  The log of the observations is 
shown in Table~\ref{LogObsTable}.

   \begin{table*}
      \caption[Target list]{Target list.}
         \label{TargetListTable}
     $$ 
         \begin{tabular}{llllllllll}
            \hline\hline
            \noalign{\smallskip}
            OBJECT                  & RA$^\mathrm{a}$ &
            DEC$^\mathrm{a}$ & L (\lsuno)& Ref$^\mathrm{b}$ & D (kpc)& Ref$^\mathrm{c}$ & \mj$^\mathrm{d}$
            & \mh$^\mathrm{d}$  & \mk$^\mathrm{d}$  \\
            \noalign{\smallskip}
            \hline
            \noalign{\smallskip}
            \glftn     & 03 01 31.3 & +60 29 13 & \eng{2.1}{4}  & 15,17& 2.20  & 16   & 12.9 &10.0  & 7.7  \\  
            \glfts     & 03 07 24.6 & +58 30 44 & \eng{3.0}{4}  & 18   & 2.70  & 5    & 12.9 &11.3  & 9.8  \\ 
            \glf       & 03 27 38.8 & +58 46 60 & \eng{1.4}{3}  & 7    & 0.90  & 19   & 10.0 & 7.6  & 5.3  \\ 
            \lkhao     & 04 30 14.4 & +35 16 25 & \eng{1.2}{4}  & 7    & 0.80  & 6    &  8.1 & 5.7  & 3.1  \\ 
            \sfour     & 05 40 52.3 & +35 41 31 & \eng{1.2}{3}  & 10   & 1.80  & 9    & 11.0 & 9.9  & 9.0  \\ 
            \sone      & 05 41 06.9 & +35 49 39 & \eng{1.5}{3}  & 10,11& 1.80  & 9    & 12.2 &10.2  & 8.1  \\ 
            \sttft     & 05 41 11.0 & +35 50 02 & \eng{2.5}{3}  & 10   & 1.80  & 9    & 12.2 &8.8   & 6.5  \\
            \ngctwo    & 05 41 45.8 & -01 54 29 & \eng{1.1}{4}  & 21   & 0.40  & 20   & 11.6 &7.4   & 4.6  \\   
            \mont      & 06 07 45.8 & -06 22 54 & \eng{5.0}{3}  & 22   & 0.95  & 1,2  & 18.0 &14.1  & 9.5  \\ 
            \mon       & 06 07 48.1 & -06 22 55 & \eng{6.0}{3}  & 1    & 0.95  & 1,2  & 12.7 & 9.5  & 6.6  \\ 
            \gfive     & 06 08 53.4 & +21 38 29 & \eng{1.1}{4}  & 23   & 1.50  & 23   & 11.6 & 10.7 & 10.3 \\ 
            \stwo      & 06 12 53.9 & +17 59 23 & \eng{3.2}{4}  & 25   & 2.40  & 24   & 16.5 & 12.9 & 9.6  \\ 
            \stwot     & 06 12 54.0 & +17 59 23 & \eng{3.2}{4}  & 25   & 2.40  & 24   & 11.3 & 10.6 & 10.4 \\ 
            \gla-W     & 06 34 37.4 & +04 12 43 & \eng{6.3}{3}  & 26   & 1.60  & 8    & 10.4 & 8.9  & 7.7  \\ 
            \gla-E     & 06 34 37.7 & +04 12 44 & \eng{1.5}{3}  & 26   & 1.60  & 8    & 14.6 & 10.5 & 7.8  \\ 
            \fan       & 06 34 35.6 & +04 12 46 & --            &      & 1.60  & 8    & 13.1 & 11.4 & 9.7  \\ 
            \glb       & 06 41 09.7 & +09 29 36 & \eng{4.0}{3}  & 27   & 0.75  & 14   & 11.0 & 7.6  & 4.8  \\ 
            \sir       & 20 27 26.8 & +37 22 48 & \eng{2.0}{4}  & 29   & 0.60  & 28   & 10.4 & 7.7  & 5.9  \\ 
            \glt       & 20 29 24.9 & +40 11 20 & \eng{4.0}{4}  & 30   & 1.50  & 30   & 14.4 &11.0  & 6.1  \\ 
            \sonef     & 22 19 18.1 & +63 18 47 & \eng{5.0}{3}  & 3    & 0.90  & 4    & 11.7 & 8.6  & 6.2  \\
            \ngcsfte   & 23 13 45.4 & +61 28 12 & \eng{5.0}{4}  & 12   & 2.70  & 13   & 15.3 &14.1  & 8.9  \\ 
          \noalign{\smallskip}
            \hline
         \end{tabular}		  
     $$ 
      \begin{list}{}{}
      \item[$^{\mathrm{a}}$] Equinox J2000.
      \item[$^{\mathrm{b}}$] References for the luminosity.
      \item[$^{\mathrm{c}}$] References	for the distance.
      \item[$^{\mathrm{d}}$] Magnitudes extracted from the literature.
      \item[Key to the references:] 1. \citet{Beckwith76};
       2. \citet{Thronson80a} ; 3. \citet{Mozurkewich86}	;
     4. \citet{Crampton74} ; 5. \citet{CohenKuhi77} ;
     6. \citet{Herbig71} ; 7. \citet{Harvey79a} ; 8. \citet{Blitz80} ;
     9. \citet{EvansBlair81} ; 10. \citet{Evans81} ;
     11. \citet{Nordh84} ; 12. \citet{CampbellThompson84} ; 
     13. \citet{Blitz82} ; 14. \citet{Park00} ; 15. \citet{Snell88} ;
     16. \citet{BeckerFenkart71} ; 17. \citet{Thronson80b} ;
     18. \citet{Wynn-Williams81} ; 19. \citet{Harvey79b} ;
     20. \citet{AnthonyTwarog82} ; 21. \citet{Jiang84} ;
     22. \citet{Howard94} ; 23. \citet{Wu96} ; 24. \citet{Itoh01} ;
     25. \citet{Howard97} ; 26. \citet{Castelaz85} ;
     27. \citet{Harvey77} ; 28. \citet{Staude82} ;
     29. \citet{Harvey82} ; 30. \citet{WynnWilliams82}
      \end{list}
   \end{table*}
%

   \begin{table*}
      \caption[Observational information]{Observational Information.}
      \label{LogObsTable}
     $$ 
      \begin{tabular}{lllllll}
         \hline\hline
         \noalign{\smallskip}
         OBJECT       & Filter /                 & Standard  & Average                & \texpo & Number  & Notes$^{\mathrm{e}}$ \\ 
                      & Telescope$^{\mathrm{a}}$ &   Star    & Counts$^{\mathrm{b}}$  &        & Frames  & \\ 
         \noalign{\smallskip}
         \hline
         \noalign{\smallskip}
         \glftn       & K / C &  SAO~12585 / SAO~23728   & 120    & 0.20 & 1536 & N! \\ 
         \glftss      & K / U &  CMC~302423 / CMC~601567 &  21    & 0.13 & 1200 & N \\ 
         \glf         & J / U &  CMC~401688 / CMC~302688 &  31    & 0.20 & 1500 & -- \\ 
                      & H / U &  SAO~24017 / SAO~23967   & 116    & 0.05 & 4000 & N! \\ 
                      & K / U &  SAO~24017               & 474    & 0.02 & 3500 & U \\ 
                      & \brq / U & SAO~24017             & 107    & 0.13 & 2000 & U \\ 
                      & Con / U &  SAO~24017             &  70    & 0.13 & 2000 & U \\ 
                      & H / C &  SAO~23999$^\mathrm{c}$ / SAO~24017 & 120  & 0.06 & 1280 & N \\ 
                      & K / C &  SAO~24098 / SAO~24017   & 537    & 0.06 & 1000 & U \\ 
         \lkhao\      & J / U &  SAO~57254 / SAO~57264   &  73    & 0.05 & 3000 & C \\ 
                      & H / C &  SAO~57239 / SAO~57264   & 692    & 0.06 & 1536 & C \\ 
         \sfour       & H / U &  SAO~58357 / SAO~58383   &  65    & 0.13 & 3000 & U \\ 
                      & K / C &  SAO~58343 / SAO~58272   &  54    & 0.06 & 1536 & U \\ 
         \sone\       & H / UC&  SAO~58357               &  56    & 0.20 & 2000 & N \\ 
                      & K / U &  SAO~58343               & 119    & 0.05 & 2000 & N \\ 
                      & K / C &  SAO~58343 / SAO~58272   & 122    & 0.20 & 1536 & N \\ 
         \sttft       & K / C &  SAO~58343 / SAO~58272   & 190    & 0.06 & 1536 & U \\ 
                      & H / C &  SAO~58357               & 107    & 0.10 & 1536 & U \\ 
         \ngctwo      & K$^\mathrm{d}$ / C &  SAO~132480 & 853    & 0.06 & 1536 & U \\ 
                      & H / C &  SAO~132410 / SAO~132434 & 144    & 0.06 & 1536 & U \\ 
         \mont        & K / U &  SAO~132881 / SAO~132874 &  15    & 0.20 & 1536 & U \\ 
         \mon         & K / C &  SAO~133884 / SAO~132874$^{c}$ & 106 & 0.06 & 1536 & N \\ 
                      & H / U &  SAO~132874              & 139    & 0.20 & 2000 & N \\ 
                      & K / U &  SAO~132884              &  75    & 0.02 & 2000 & N \\ 
         \gfive       & K / C &  SAO~78056$^{\mathrm{c}}$ / SAO~77997& 23  & 0.3 & 768  & U \\ 
         \stwo/3      & K / C &  SAO~95350               &  41    & 0.20 & 1536 & N \\ 
         \gla -E/W    & H / U &  SAO~114047 / SAO~114O94 & 209    & 0.25 & 1000 & N!/C \\
                      & K / C &  SAO~114047 / SAO~114O80 &  62    & 0.06 & 1792 & U \\
         \fan         & K / U &  SAO~114047 / SAO~114O94 &  88    & 0.13 & 2000 & U \\
         \glb         & H / U &  SAO~95987               & 140    & 0.05 & 3000 & U \\ 
                      & H / C &  SAO~95987               & 120    & 0.06 & 2048 & N! \\ 
                      & K / C &  SAO~114234 / SAO~114245 &  68    & 0.06 & 1792 & U \\ 
         \sir         & K / C &  SAO~76077 / SAO~70120   & 330    & 0.06 & 2048 & U \\ 
                      & H / C &  SAO~70120 / SAO~70129   & 109    & 0.13 & 1280 & U \\ 
         \glt         & H / U &  SAO~70099 / SAO~49777   &  29    & 0.13 & 2000 & -- \\ 
                      & K / U &  SAO~70052               & 381    & 0.05 & 1500 & N! \\
                      & K / C &  SAO~70052 / SAO~49730   & 274    & 0.06 & 2048 & N! \\
         \sonef       & K / C &  SAO~20051 / SAO~19948   & 250    & 0.06 & 1600 & N \\ 
         \ngcsfte     & K / C &  SAO~20455 / SAO~20563   &  21    & 0.20 & 1536 & N! \\
         \noalign{\smallskip}
         \hline
      \end{tabular}
     $$ 
      \begin{list}{}{}
      \item[$^{\mathrm{a}}$] U: UKIRT. C: 3.5~m telescope at Calar Alto.
      \item[$^{\mathrm{b}}$] Average peak counts per frame.
      \item[$^{\mathrm{c}}$] These stars were found to be close binaries and so not included in the reduction.
      \item[$^{\mathrm{d}}$] Object was too bright at K and so a narrow-band filter at 2.26$\mu$m was used instead.
      \item[$^{\mathrm{e}}$] Main features observed in the reconstructed images. N:
	Bright nebula likely to be associated to the outflow cavity. N!:
	Very low level or hardly resolved nebulosity. C: Close companion
	(with a separation $<$\,1\farcs5). U: Unresolved source. --:
	Reconstruction not possible due to the low number of
	photons per individual frame. 
      \end{list}
   \end{table*}

The images were carefully cleaned of bad pixels, since they can
contaminate the visibility function with bright peaks at specific
spatial frequencies. An average sky image (produced by averaging all
sky frames corresponding to one target) was subtracted from every
snapshot of the object and reference star. Cleaned and sky subtracted
images were Fourier transformed and used to calculate the power
spectrum and the bispectrum \citep[e.g.][]{Hofmann86}. The
visibilities for each target were calculated using the
corresponding reference star listed in Table~\ref{LogObsTable}. The
phases were recovered with a least-squares method using 80 bispectral
planes in the \fovp{128}{128}\ images and 28 bispectral planes in the
\fovp{256}{256}\ images \citep{Glindemann91}. The visibilities were 
filtered using a low-pass Gaussian window to reduce the noise level at 
high spatial
frequencies. Real images are obtained from the inverse Fourier
Transform of the modulated visibility and the phase. The modulation of
the visibility results in the resolution of the final images being
lower than the full diffraction limit of the telescope, but greatly
improves the detection level of faint extended nebulosity. Generally,
the reconstructed image shows artifacts, which can be the result of
the loss of part of the speckle cloud off the edge of the detector, or
due to a poor phase recovery. They usually appear concentrated close
to a strong peak, and are at the level of a few percent of the
brightest peak. Tests on reference stars were used to check for the
level of these artifacts, as well as for binarity amongst the
reference stars themselves.

\subsection{Calar Alto Observations}

The observations were made at the 3.5m telescope on Calar Alto, on
December 9, 10 and 11, 1994. The MAGIC 256$\times$256 NICMOS 3 camera
was used at a pixel scale of 0\farcs073 per pixel in order to sample 
the diffraction limit at
K. For simple point sources 1500 128$\times$256 frames with an
integration time short enough to freeze the seeing were accumulated on
the object interspersed with a similar number on one or more reference
stars less than 1$^\circ$\ away. The seeing was approximately 1 arcsec. 
Only half of the array was saved in order to save disk space and speed 
up the readout. The source was alternately placed in lower and upper 
quadrants, the blank one being used for sky subtraction. Where
multiple sources or complex structure extended across the field of
view the whole array was saved and separate sky frames where
obtained. The reduction procedure is similar to the one used for the
UKIRT images.

A typical value for the limiting surface brightness in 
our observations can be obtained from the seeing-limited magnitudes
listed in Table~\ref{TargetListTable} and the typical values for the
resolution and noise level in the restored images (see 
figures~\ref{KMonR2SpeckleFig} to \ref{GL5180peckleFig}).
As an illustration, we take the instance of \glf\ in the H band 
from our Calar Alto campaign, since the reconstructed image
(see Fig.~\ref{HGL490CAFig}) is dominated by a strong point source 
with only a faint nebulosity and no other sources present in the 
field of view. With a seeing-limited magnitude of 7.6 and a 
restored resolution of 0\farcs23, only 8\% of the flux is  
in the central pixel. A 1$\sigma$\ noise level of 2.5\% of 
the brightness peak is found within 1$''$\ of the peak, whilst 
$\la$\,1\% noise is found further out. These yield 3$\sigma$\ 
limiting surface brightnesses of 7.4~mag/arcsec$^2$\ and 
10.1~mag/arcsec$^2$\ respectively. 
These estimates are similar for the rest of the sources
for which we could reconstruct a high resolution image. Therefore, 
the limiting surface brightness (in units of mag/arcsec$^2$) in 
the region of interest (i.e. within 1$''$\ of the central peak) is 
of the same order as the total seeing magnitude (in units of
mag). This is less valid for sources with a strong nebulosity or 
for fields with multiple sources. The noise level is determined by the
dynamic range ($\sim$\,100) of the speckle technique used here.

\section[Results]{Results}
\label{ResultsSection}

In this section we present the results on individual objects before
discussing the sample as a whole in the next section.

\subsection[Mon\,R2\,IRS3]{\mon}

\mon\ is the brightest of a cluster of compact IR sources with
associated extended emission located at a distance of 950~pc.
IRS3 is located approximately
$30''$~to the east of a blister type \hii\ region of $\sim
27''$~diameter \citep{Massi85}.  The cluster is located near the
centre of a giant CO outflow \citep{BallyLada83,Wolf90}
oriented at a \pa{135} with the approaching gas flowing to the NW.
Due to the low resolution of the CO observations, 
it is not clear which of the IR sources is powering this outflow. On scales
of a few arcseconds, the IR reflection nebula is extended to the south
and to the east.  The polarisation pattern shows centro-symmetry
indicating that IRS3 is the illuminating source. It also shows a
polarisation disc elongated in the SE-NW direction \citep{Yao97}.

\begin{figure}
  \centering
  \resizebox{\hsize}{!}{\includegraphics{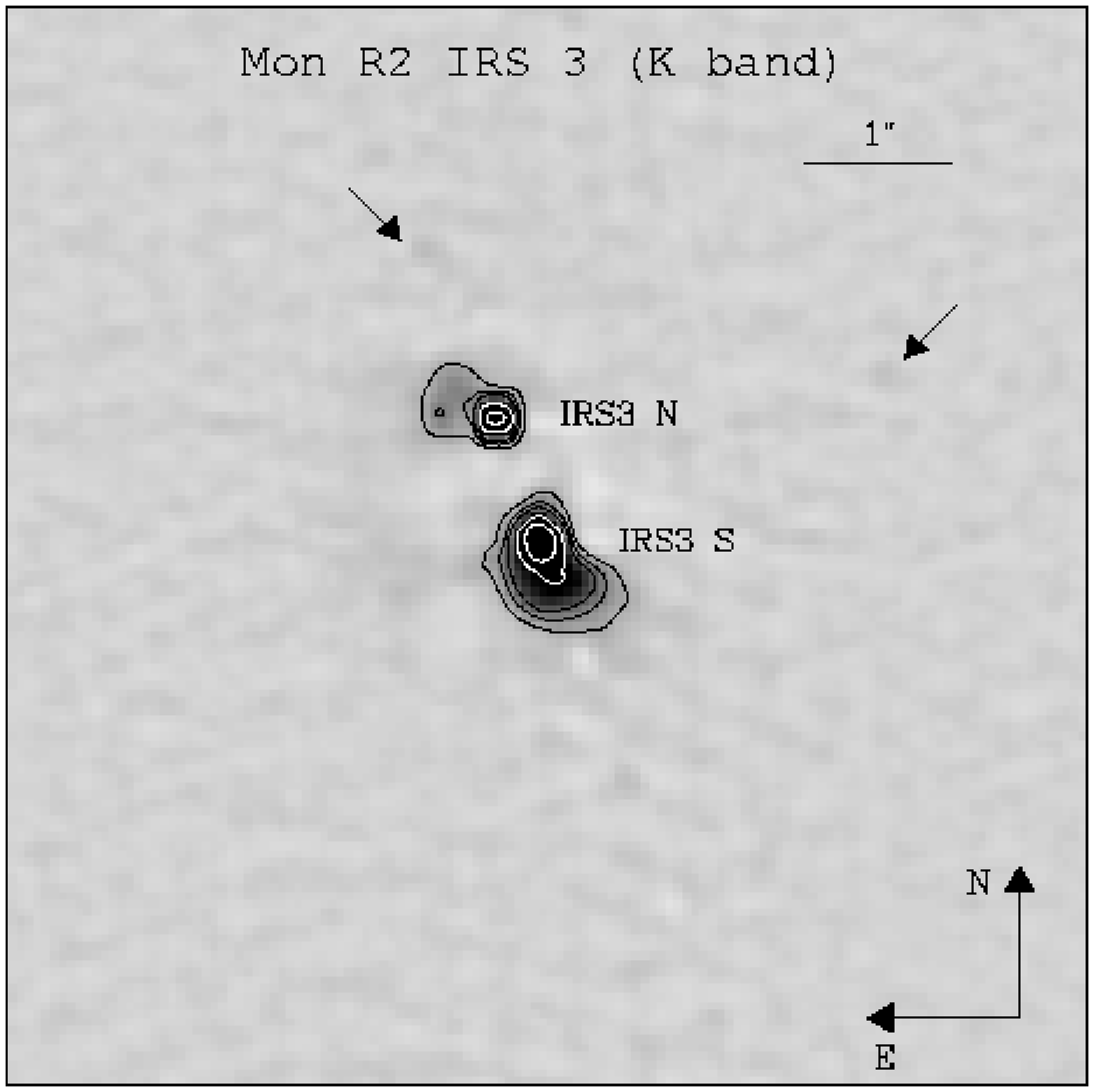}}
  \caption[Reconstructed image of \mon\ in the K band]{Reconstructed
    image of \mon\ in the K band. A Gaussian window of
    FWHM=\sfreq{5.3}\ was applied to the visibility to achieve a
    resolution of 0\farcs19. The contours are at 
    5, 10, 15, 30 and 50~\% of the peak. The noise level is
    \anyapp{0.4}{\%}. The sources are labelled using the notation by
    Koresko et al. 1993. The arrows indicate faint stars used to align the
    H and K images. In this and in the following
    reconstructed images the intensity scale is normalised to the
    brightest peak with the brighter regions represented by the darker
    greys.}
  \label{KMonR2SpeckleFig}
\end{figure}

\begin{figure}
  \centering
  \resizebox{\hsize}{!}{\includegraphics{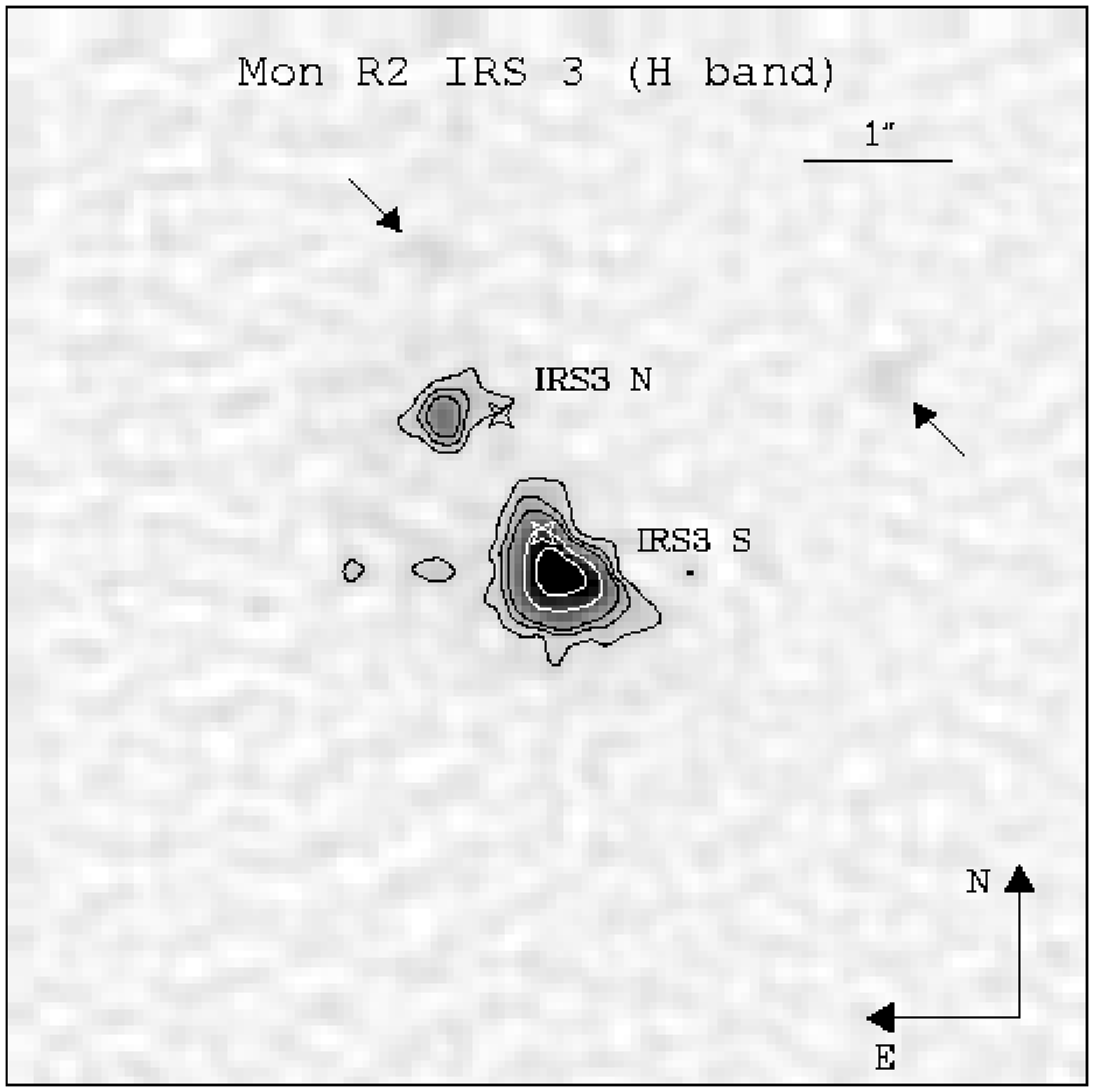}}
  \caption[reconstructed image of \mon\ in the H band]{Reconstructed image of 
    \mon\ in the H band. The resolution is 0\farcs19. The contours are at 
    5, 10, 15, 30 and 50~\% of the peak and the noise level is \anyapp{1}{\%}. 
    The labels used are the same as for the K band image. The star 
    symbols indicate the position of the stars IRS3\,N and IRS3\,S in the 
    K~band image. The small arrows indicate the position of the two stars 
    used to align this image and the K band image.}
  \label{HMonR2SpeckleFig}
\end{figure}

\begin{figure}
  \centering
  \resizebox{\hsize}{!}{\includegraphics{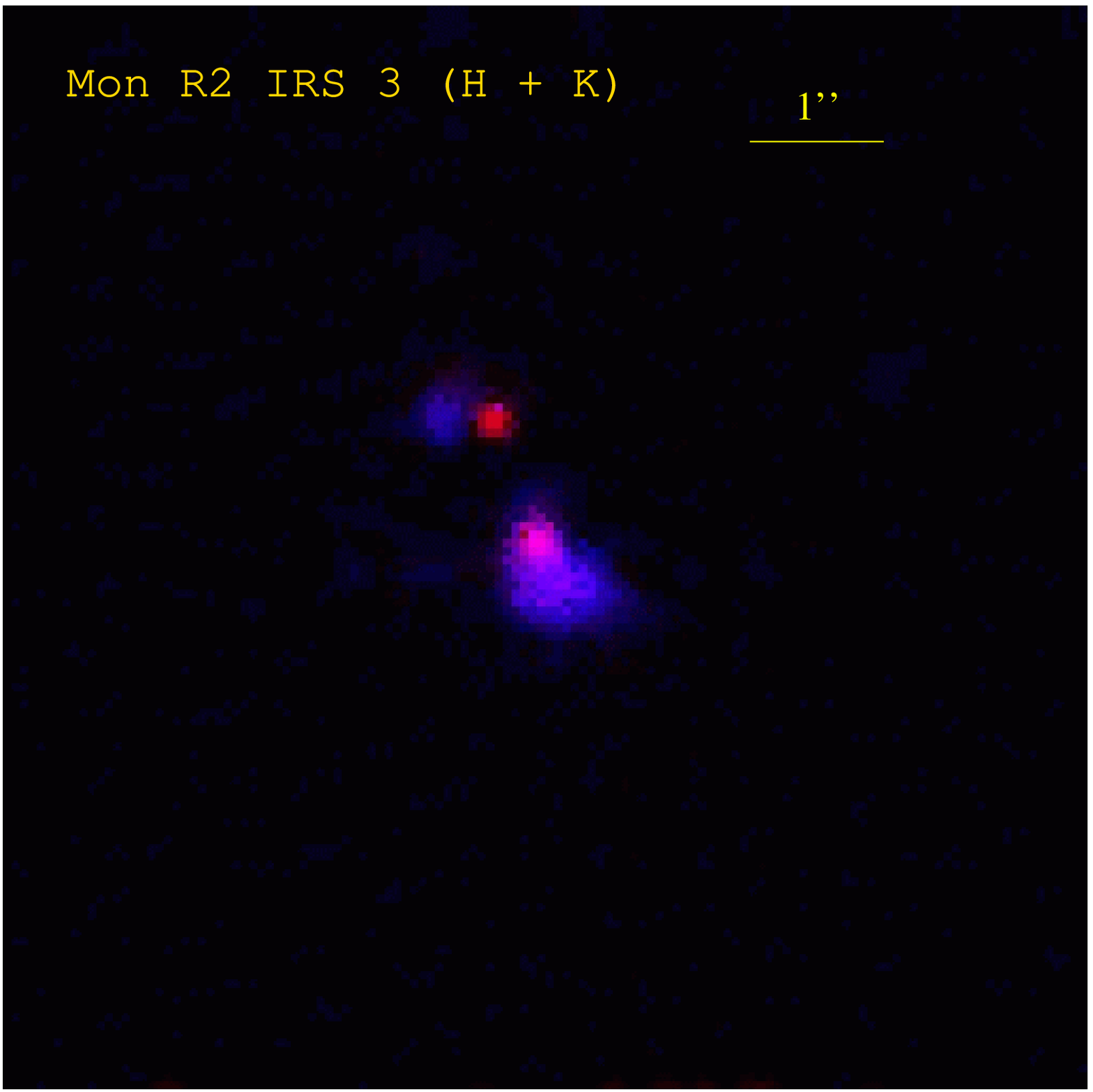}}
  \label{MonR2CoFig}
  \caption[Colour-coded image of \mon]{Colour-coded image of \mon. The 
    emission at H and K are represented by the blue and red colours 
    respectively. The blue colour of the nebulae indicates the 
    scattered nature of the light.}
\end{figure}

This object has been investigated with speckle imaging before by
\citet{Koresko93} at 2.2, 3.8 and \mum{4.8} and more recently in deep
images at H and K by \citet{Preibisch02} using the 6 m SAO
telescope.  We present our K and H-band speckle images in
figures~\ref{KMonR2SpeckleFig} and~\ref{HMonR2SpeckleFig} for
comparison with the rest of our survey. In the K-band two stars are
clearly present, IRS3\,S and IRS3\,N, that are separated by
\anypm{0\farcs87}{0\farcs03}, in good agreement with the works of
\citet{McCarthy82}, \citet{Koresko93} and \citet{Preibisch02}. 
IRS3\,N is actually composed of two stars (sources B and C in
  \citealp{Preibisch02}). These two stars are also seen by Koresko 
  et al. in the K band. However, they are only marginally resolved in our 
  K band data.
The conical reflection nebula emanating to the S-SE
of IRS3\,S found by \citet{Koresko93} is clearly visible in both the H
and K bands. At H, 
the brightest of the two stars in IRS3\,N (source B in the work by 
  Preibisch et al.) disappears completely, whilst the
star illuminating the southern nebula is faint, if present at all, at
the northern tip of the conical nebula. These monopolar nebulae are
interpreted as light scattered off the walls of an outflow cavity that
is directed somewhat towards us. The other lobe that would be
associated with the redshifted outflow is not visible due to high
extinction in an inferred flattened envelope perpendicular to the
bipolar outflow direction. Figure~\ref{MonR2CoFig}, represents a
colour-coded image, which stress the presence of the highly reddened 
stars and the relatively blue conical nebula
associated with IRS3\,S. The scattering efficiency of dust increases
rapidly with decreasing wavelength through the near-IR, causing the
reflection nebula to be relatively blue.  

The morphology of the southern nebula indicates that in this case the 
outflow is at a position angle (PA) of \angpm{198}{15} with the
approaching gas 
towards the south-west. \citet{Preibisch02} show that the emission to
the E of IRS3\,N is the combination of another star and a jet-like
feature emerging from IRS3\,N at a \pa{50}. Neither of these outflow
orientations correlate with the large scale molecular outflow 
described by \citet{Wolf90} (\pa{135}\, with the blue gas moving towards 
the north-west). Therefore, the IRS3 sources are not likely to be
driving the large scale bipolar molecular outflow in the region and we 
need to look for other outflow indicators nearer to IRS3 itself.
One should keep in mind though, that single-dish CO observations 
  have a very low spatial resolution (several arcseconds) compared
  with our speckle images. The giant molecular outflow in the Mon\,R2 
  region maybe a
  superposition of several smaller outflows with slightly different 
  orientations, each one driven by a different source. Therefore, any
  attempt to relate the near-IR sub-arcsecond morphology with large 
  scale structures should be taken with caution, even though this is 
  an interesting exercise to search for possible correlations in 
  orientation.

It is also interesting to compare the orientation of the cavity with the 
distribution of methanol masers presented by \citet{Minier00}.
They show a group of masers forming a linear structure with an approximate 
length of 0\farcs17 at PA\,=\,\angpm{230}{20}. They interpret this structure in 
terms of a nearly edge-on circumstellar disc. This would imply a 
system with a rotation axis in the NW-SE direction, which is in 
complete disagreement with what the near-IR sub-arcsecond nebula suggests. 
Alternatively, the situation maybe similar to other sources (e.g. W75N
\citealp{Torrelles97}) where the linear maser morphology appears to trace 
the outflow axis.

Finally, the orientation of the polarisation structure shown by 
\citet{Yao97} is considered. Their seeing-limited images indicate the 
presence of a polarisation disc at \pa{140}, which is roughly perpendicular 
to the maser structure. It is likely that IRS3\,S, which appears to be 
the strongest source of scattered light in our UKIRT speckle images, is 
the main contributor to the polarisation disc. However, high resolution 
polarimetry in the near-IR is necessary to resolve the polarisation 
structure of each of the components independently.

\subsection[S140\,IRS1]{\sonef}


\sonef\ is located near the centre of a
bipolar molecular outflow at a \pa{160}\ with the blue-shifted lobe
pointing towards the SW
\citep{Blair78,BallyLada83,Hayashi87,Minchin93}. 
\citet{Minchin95} find that the \cseveno\ emission at the ambient
cloud velocity traces the cavity wall associated with the blue-shifted
CO outflow lobe.

Seeing-limited imaging polarimetry in the optical and in the IR show
that \sonef\ is the brightest member of a cluster of embedded sources
surrounded by an extended reflection nebula
\citep{ForrestShure86,Lenzen87,Evans89,Harker97,Yao98}. The
polarisation structure of the nebula indicates that IRS1 and IRS3 are
the main illuminating sources of the nebula, but the polarisation
structure is complex.

Radio continuum observations reveal a thermal source with a spectral
index of $0.8$\ indicative of an ionised wind. Observations that
resolve the source show that the radio emission is elongated {\em
perpendicular} to the large scale outflow direction \citep{Schwartz89,
Tofani95}.  The higher resolution observations by
\citet{HoareMuxlow96} show a structure at a PA of
\angpm{44}{1}. Multi-epoch observations show that this structure is
not a radio jet as first proposed by \citet{Schwartz89}, but an
equatorial wind \citep{Hoare02}, as was also found for S106\,IR
\citep{Hoare94} and possibly GL\,490 \citep{Campbell86}.

\begin{figure}
  \centering
  \resizebox{\hsize}{!}{\includegraphics{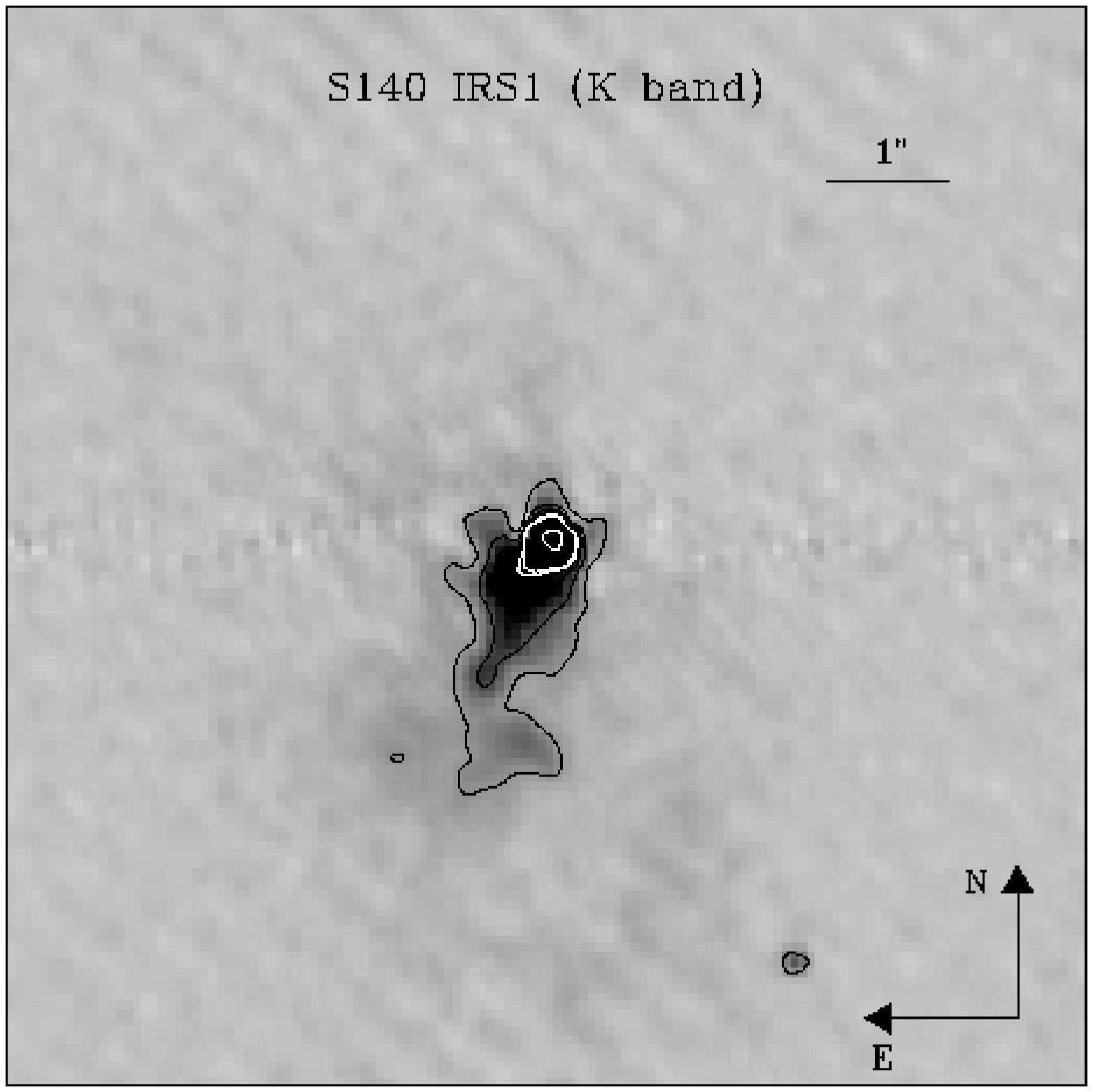}}
  \caption[Reconstructed image of \sonef\ in the K~band]{Speckle 
    reconstructed image of \sonef\ in the K~band at a
    resolution of 0\farcs19. The contour levels are at 1\%, 2.5\%,
    4\%, 5.5\%, 7\%, 8.5\% and 50\% of the peak.  The noise
    level is $<1\%$. The grey-scale varies from -2\% (white)
    to 8\% (black).}
  \label{S140SpeckleFig}
\end{figure}

Our K-band speckle image of \sonef\ is shown in 
figure~\ref{S140SpeckleFig} (see also \citealp{Hoare96}). This
image shows a monopolar nebula that emerges from IRS1 and extends 
towards the SE at a \pa{150}$\pm15^\circ$. 
The azimuthally averaged visibility function indicates
that \anypm{22}{2}\% of the flux arises from the point source, which
is likely to be the central proto-star \citep[see ][]{Schertl00}. 
The morphology in figure~\ref{S140SpeckleFig} agrees with deeper
speckle imaging by \citet{Schertl00}. Their polarisation 
data also show the centro-symmetric pattern that confirms
that this is a reflection nebula illuminated by \sonef.
This is consistent with the light being reflected off the walls of 
the blue-shifted outflow cavity in a classic monopolar reflection 
nebula. All this fits in with the interpretation of the radio 
structure as an equatorial wind.

The only question mark over this picture is the discovery by 
\citet{Weigelt02} of a set of three bow shock-like structures 
to the NE of \sonef\ in wide-field K-band speckle images.
They interpret these as arising from precessing, episodic jet activity
from \sonef. This would require a return to the jet interpretation 
for the radio emission, which is inconsistent with the direction of
the main outflow indicators, i.e. the CO outflow and reflection 
nebula. One solution to this conflict would be if the bow shock
structures were driven by a jet from a lower-mass YSO located near 
to \sonef.

\subsection[S255\,IRS1 and IRS3]{S255\,IRS1 and IRS3}

\stwo\ and IRS3 form part of a cluster of IR sources located between
the evolved \hii\ regions S255 and S257 at a distance of 2.5~kpc.  A
molecular outflow along the N-S direction with its approaching lobe
towards the south appears to be centred near the IR cluster
\citep{Heyer89}. Infrared polarisation images by \citet{Tamura91} show
a bipolar nebula (IRN1) at PA of \degg{63} illuminated by IRS3, which
indicates the presence of an outflow in the this direction. They also
find another reflection nebula (IRN2) to the north of IRS1 that is
illuminated by that source. Thermal IR observations by \citet{Itoh01}
show that IRS3 is the more luminous and embedded object.
\citet{Miralles97} find molecular hydrogen 
emission from HH-like objects along the jet direction.

\begin{figure}
  \centering
  \resizebox{\hsize}{!}{\includegraphics{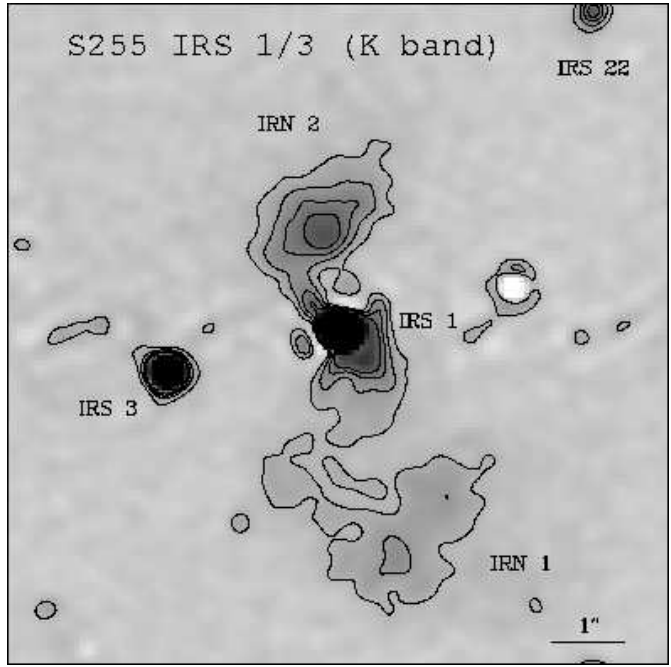}}
  \caption[Reconstructed image of \stwo\ in the K~band]{Speckle 
    reconstructed image of \stwo\ in the K~band at a
    resolution of 0\farcs29. The contour levels are at 0.5\%, 1\%,
    1.5\% and 2\% of the peak.  The noise level is $<1\%$. The 
    grey-scale varies from -1\% (white) to 4\% (black). The
    negative feature (-3.5\%) west of IRS1 is a 'ghost' of IRS3.}
  \label{S255SpeckleFig}
\end{figure}

The K-band reconstructed image of this region is shown in
figure~\ref{S255SpeckleFig}. The field 
of view contains the point sources IRS1, IRS3, IRS22 and the two reflection
nebulosities IRN1 and IRN2 using the nomenclature of \citet{Tamura91}.  
IRN1 is seen to be very conical in shape
with an opening angle of \degg{35}\ and at \pa{58} from its illuminating
source IRS3. Although the
main orientation of the CO outflow is along the N-S direction, 
the CO observations of \citet{Heyer89} show some evidence for high 
velocity blue-shifted gas towards the SW roughly along the outflow axis
indicated by IRN1 in figure~\ref{S255SpeckleFig}.

IRN2, which Tamura et al. deduced was present from the polarisation
pattern, is clearly resolved in our observations. There is a gap
between the main IRN2 nebula and IRS1 and its morphology is not
entirely consistent with IRS1 being its illuminating source.
The presence of foreground clumps or intra-cluster matter 
could be a possible explanation for this gap.
Significant artifacts exist on the RA axis through IRS1 so that the
reality of features close to IRS1 are questionable.  Some spatially
distinct diffuse emission is seen on the opposite side of IRS1, which
maybe the other reflection lobe. If these nebulosities were associated
with the main N-S outflow, we would expect the southern blueshifted
lobe to be brighter. 

\subsection[GL\,437-S]{GL\,437-S}

\glfts\ is a compact cluster of IR sources \citep{Wynn-Williams81} located a 
distance of 2.7~kpc. The cluster, with a total
luminosity of $\sim$~\lexpsun{3}{4}, comprises four bright sources 
denominated \glftsn, S,~E and~W. 
The cluster is located at the centre of a poorly collimated molecular outflow 
\citep{Gomez92}. \citet{WeintraubKastner96} find that the cluster is embedded 
in an infrared polarised nebula. The polarisation pattern is mainly 
centro-symmetric with respect to the embedded source WK~34, which is close to 
\glftsn. However, centro-symmetry can also be observed about \glftss\ 
\citep{Weintraub96}.

\begin{figure}
  \centering
  \resizebox{\hsize}{!}{\includegraphics{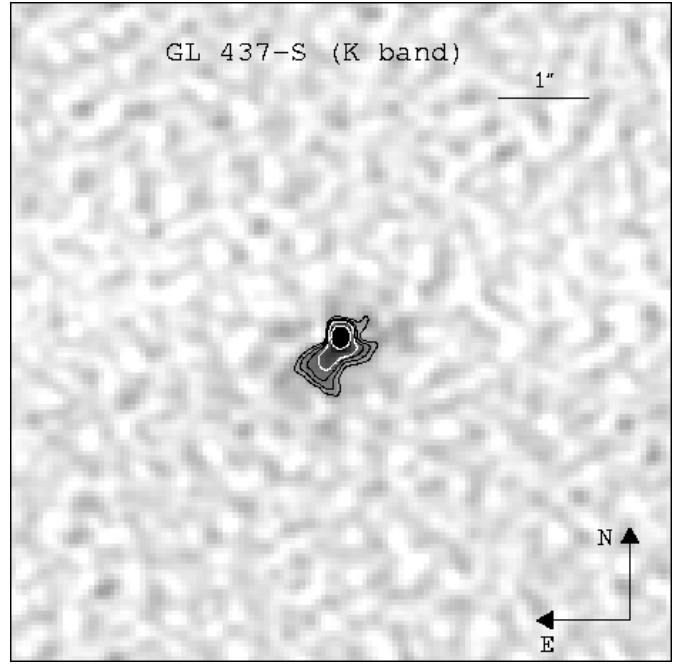}}
  \caption[Speckle reconstructed image of \glftss\ in the K~band]{Speckle 
    reconstructed image of \glftss\ in the K~band. The resolution
    attained is 0\farcs22. The colour scale varies from -1\% (white) to 5\% 
    (black) of the brightness peak. The contours are at 15\%, 20\%, 25\%, 
    30\% and 50\% of the peak, and the noise is at the 4\% level.}
  \label{KGL437SpeckleFig}
\end{figure}
 
Figure~\ref{KGL437SpeckleFig} shows the reconstructed image of \glftss\ 
at a resolution of 0\farcs22. The data indicate the presence of a 
sub-arcsecond monopolar nebula extended towards the SE at a 
PA\,=\,\anypm{135^\circ}{10^\circ}. The noise level is high in this
image since \glftss\ is quite faint, but the extended emission is 
bright relative to the star. However, the presence of the nebulosity
is consistent with the centro-symmetry in the near-IR polarisation 
seen by \citep{Weintraub96}. 
The high-velocity CO emission shows several peaks (see figure~3 of 
Gomez et al. 1992). Although the main outflow axis appears to be in the N-S 
direction, the presence of multiple outflows in the region cannot be 
discarded. In particular, there is some blue-shifted emission towards the~SE, 
and red-shifted gas to the NW, that may trace a bipolar outflow oriented 
roughly along the same direction as the sub-arcsecond nebula shown in 
figure~\ref{KGL437SpeckleFig}. 

\subsection[LkH$\alpha$\,101]{\lkhao}

\lkhao\ is a Herbig~Be star with strong \hal\ emission \citep{Herbig56} which 
illuminates the reflection nebula NGC~1579 \citep{Redman86}.
It is the brightest of a cluster of IR sources \citep{Barsony91} at a distance 
of 800~pc. Its spectral type corresponds to a late~O or early~B 
star \citep{Brown76,Harvey79a}. It excites a weak \hii\ region of size 
$\sim 1'$\ 
\citep{Becker88}. \lkhao\ is also the source of a compact ionised wind. The
radio continuum emission from the wind has been studied at a wide range of 
frequencies \citep{Brown76,Bieging84,Becker88} showing
a  $\sim \nu^{0.6}$\ law. \citet{HoareGarrington95} resolve the wind
into a clumpy and approximately circular distribution on the sky. 

\citet{BallyLada83} studied the large scale molecular gas motions in the 
region. They classified \lkhao\ as an intermediate velocity source (with 
velocities in the 10~-~30~\kmsc\ range). However, the existence
of a bipolar molecular outflow was not clear from their data, and 
has been completely discarded by \citet{Barsony90}. This object shares many 
of the common features of massive young stellar objects, but the fact 
that it is seen in the optical indicates that it is probably 
in a fairly advanced evolutionary stage.

The sub-arcsecond structure in the near-IR of \lkhao\ has been studied 
by several authors using 1D~speckle interferometry
\citep{Dewarf93,Leinert97,Leinert01}.  They classify the source as
either unresolved or with a core-halo morphology. However, this object 
was resolved into two components (the brightest surrounded by a disk) 
with a separation of 0\farcs18 and at a \pa{70}\ by
\citet{Tuthill01,Tuthill02} using aperture-mask interferometry with
the Keck~I telescope at a resolution of \anyapp{0\farcs023}. 

For completeness, we show in figure~7 our J~band
speckle data of \lkhao\ taken at UKIRT at a resolution of 0\farcs11.
A separation of \anypm{0\farcs17}{0\farcs02}\ between both components was 
measured directly in the reconstructed image and also by fitting the 
bispectral phase. The position angle of the fainter component is 
\angpm{75}{5}. Aperture photometry on the reconstructed image as well
as fitting the bispectrum yielded a brightness of the fainter
component with respect to the total of \anypm{0.17}{0.03}. The 
binary system can also be seen in the H~band data taken during the 
Calar Alto campaign. A separation of \anypm{0\farcs19}{0\farcs03}\ and
a relative brightness of \anypm{0.06}{0.02} with respect to the
total. These values are consistent with those measured by Tuthill et
al. at higher resolution.

\begin{figure}
  \centering
  \resizebox{\hsize}{!}{\includegraphics{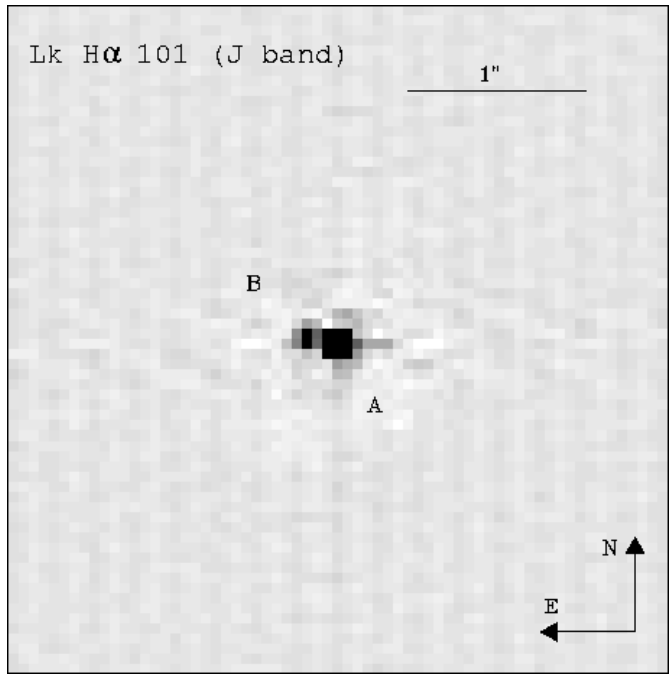}}
  \label{Ha101Fig}
  \caption [Reconstructed image of \lkhao 
    at J]{Reconstructed image of \lkhao\ in the J~band with a
    resolution of 0\farcs11. A Gaussian window of
    FWHM=\sfreq{9.1}\ was applied to the visibility
    to reduce the noise associated with the high spatial 
    frequencies. The faint component is located 0\farcs17 away 
    (\anyapp{130}{AU} at a distance of 800~pc) and it
    is $\sim 5$\ times fainter than the primary. The background
    noise is $\la 1$\% of the peak.}
\end{figure}

\subsection[GL\,490]{\glf}

This massive YSO is associated with a parsec scale molecular outflow
that was first detected by \citet{LadaHarvey81}. More detailed
structure of the outflow is given by \citet{Snell84},
\citet{Mitchell92,Mitchell95}. All observations agree in that the outflow
axis lies in the NE-SW  direction (\pa{45}) with the
blue-shifted lobe towards the south-west. The CO outflow is not highly
collimated, with a considerable overlap of the blue-shifted and 
red-shifted gas.

An stellar-like optical counterpart is located at the 
apex of a conical reflection nebula in the blueshifted outflow lobe 
\citep{Campbell86}. The light centre of the optical source was found
to shift to the SW at bluer wavelengths consistent with a scattered
rather than direct source for the optical light \citep{Campbell86}. Near-IR
imaging polarimetry of the region performed by \citet{Minchin91} shows
a cometary reflection nebula extending
for about $20''$\  towards the south-west, with its head at the location of 
\glf. This nebula shows a wider opening angle than its optical counterpart, 
and it is characterised by a centro-symmetric polarisation pattern with a 
polarisation disc at a PA of \angpm{120}{2}.

An ionised stellar wind is seen in the radio with a spectral index of~$\sim 1.2$\ 
\citep{Simon83,Campbell86,Henning90}. The radio source appears extended in
a direction perpendicular to the outflow at \ghz{15}\ \citep{Campbell86}.
At the position of the star, \citet{Kawabe84} found CS emission elongated in 
a direction roughly perpendicular to the outflow axis that can be interpreted 
as produced in a dense disc or torus of material around the IR source.  
\citet{Schreyer02} also find an elongated structure at a 
PA\,$\sim$-45$^\circ$\ in the CS (J=2-1) transition.  
Further evidence of the disc is obtained from the 2.7~mm continuum 
emission detected by \citet{Mundy88}, which is also elongated perpendicular 
to the outflow. 

The sub-arcsecond structure of \glf\ in the near-IR was first studied by
\citet{Haas92}. Their 1D speckle interferometry in the H, K and L bands 
shows that the source is formed by an unresolved core of \anyapp{0\farcs1}{} 
and an extended halo of $\sim$~1\farcs2. The halo appears relatively bluer,
slightly elongated towards SW (PA\,$\sim 125^{\circ}$), and shows a 
centro-symmetric polarisation pattern, which indicates a scattered origin for
the emission.  The same core-halo structure is found by \citet{Dewarf93} 
at \mum{3.8}. 

\begin{figure}
  \centering
  \resizebox{\hsize}{!}{\includegraphics{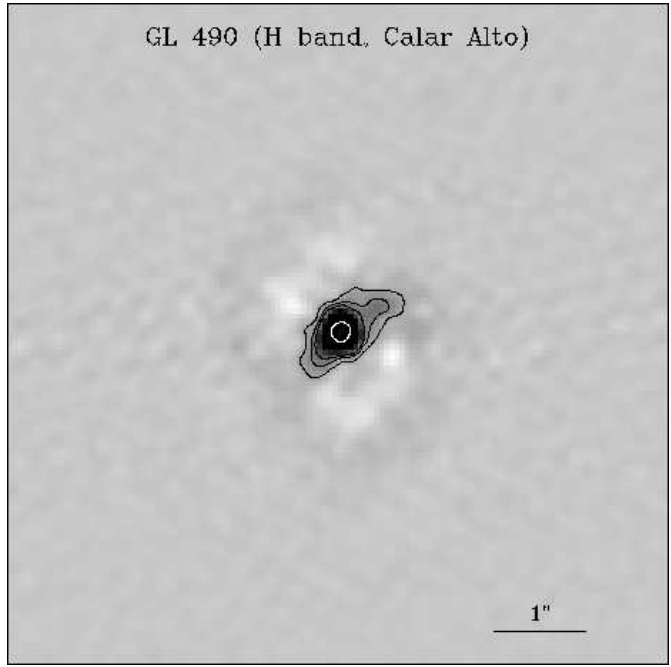}}
  \caption[H band reconstructed image of \glf\ taken at CA]{H~band
    speckle reconstructed image of \glf\ from data
    taken at Calar Alto in 1994 
    The resolution is 0\farcs23. The grey-scale
    varies between -2.5\% (white) and 9.4\% (black) of the peak. 
    The contours are at 1.5, 3, 4.0 and 50\% of the peak. The
    noise level is $0.2\%$ away from the source.}
  \label{HGL490CAFig}
\end{figure}

Our K-band speckle imaging does not resolve any sub-arcsecond
structure. However, extended emission is seen in our H~band 
reconstructed image taken at Calar Alto (see figure~\ref{HGL490CAFig}). 
This blue colour of the nebula strongly indicates that it is 
scattered light. The H-band nebulosity is mainly
perpendicular to the outflow (\pa{120}), but appears to curve off 
towards the blueshifted SW lobe at the ends. Such a morphology is
consistent with arising at the base of
the wide opening angle blueshifted outflow cavity. This structure
is very similar in size and orientation to the extended radio
emission seen by \citet{Campbell86}. 

\begin{figure}
  \centering
  \resizebox{\hsize}{!}{\includegraphics{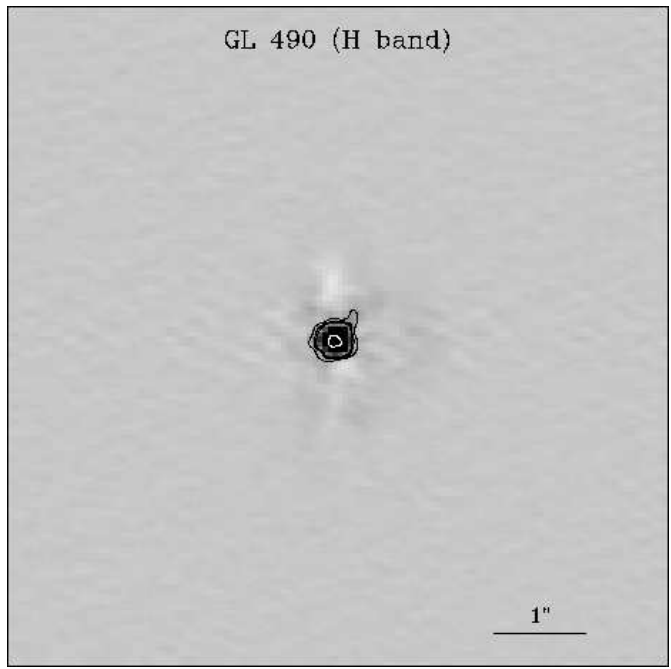}}
  \caption[H~band speckle image of \glf]{H~band speckle image of 
    \glf\ taken at UKIRT in 1997. The resolution achieved
    was 0\farcs18. The visibility was multiplied by a Gaussian 
    window with a FWHM of \sfreq{5.5}. The grey-scale varies 
    from -1\% to 3\%. The contours are at 1.5, 3, 4.5 and 30\% 
    of the peak. The most negative feature is at a -1.5\%
    level. The noise is $\sim 0.2\%$ away from the source.}
  \label{HGL490Fig}
\end{figure}

For the speckle data taken at Calar Alto (see Table~\ref{LogObsTable}), 
it turned out that one of the reference stars used
(SAO~23999) was actually a binary and so could not be used in the
reconstruction. This resulted in somewhat higher than normal 
artifacts and partly for this reason the speckle observations of
GL 490 were repeated at UKIRT. Figure~\ref{HGL490Fig} shows the H~band 
image reconstructed at a resolution of 0\farcs18 from this dataset. 
The elongated morphology seen in the image taken at Calar Alto appears 
to have vanished in the new UKIRT image. Tests done with
  cross-calibrated PSF stars favour the reality of the disappearance of 
  the nebula in GL 490.

Interestingly, the extended radio emission observed by \citet{Campbell86} 
has also not been detected in more recent deep 8 GHz VLA observations
by Hoare \& Gibb (in prep.). Hence, it is possible that these changes in the
morphology of near-IR and radio observations maybe caused by real 
variability in the source. One possible explanation
could be obscuration near the star (above the disc) preventing the
sides of the cavity being illuminated with near-IR light, thereby
reducing the reflection nebula and EUV radiation, thereby stopping
photoionisation and hence the radio emission as well. Further 
  multi-epoch observations at different wavelengths are needed to 
  probe this possible variability.

Any reflection nebulosity should be brighter relative to the star at J
compared to H. Unfortunately, GL 490 is too faint at J for  
the reconstructed image taken at UKIRT to confirm the reality of the
nebulosity. Conversely, the source is so bright at K that speckle 
imaging could be performed through a \brq\ filter. However, no
extended emission due to ionised gas was found here either.

\subsection[GL\,961 E, W and Fan]{\gla\ E, W and Fan}

\gla\ is a luminous \citep{Castelaz85} young stellar 
object that is located in the south-west rim of the Rosette nebula at a 
distance of 1.6~kpc. This source is very faint in the optical 
\citep{Eiroa81,Beetz76}, and lies approximately $30''$~E of an optical 
fan-shaped nebula (\fan). \citet{Lenzen84} found that \gla\ is actually 
composed of two close IR objects, separated by about~$5''$ (8000~AU) at a 
\pa{70}. \citet{Castelaz85} show that the spectral energy distribution of 
\glae\ dominates at wavelengths longer than \mum{2.2}, while the western 
object is brighter at shorter wavelengths. 

The double source is located at the centre of a parsec scale molecular 
outflow oriented in the N-S direction with the approaching lobe
towards the north \citep{LadaGautier82}. Observations
of the \molh\ emission in the region performed by \citet{Aspin98} show 
multiple bow-shocks emerging from the double system. 
One system is oriented at \pa{40}\, and the other one 
at \pa{0}, which corresponds with the orientation of the CO outflow.

\begin{figure}
  \centering
  \resizebox{\hsize}{!}{\includegraphics{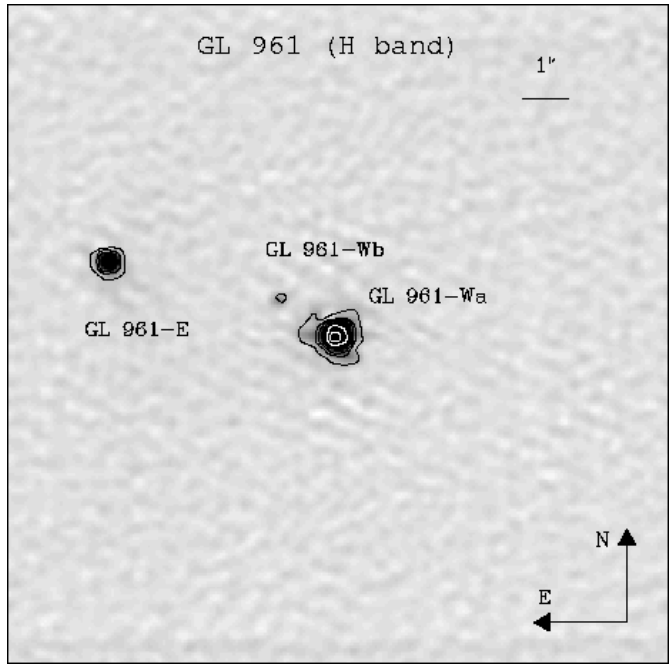}}
  \caption[Reconstructed image of \gla\ at~H] {Reconstructed
    image of \gla\ at~H. A Gaussian window with a FWHM=\sfreq{3.1}\ was 
    applied to the visibility function in order to filter out
    the noise associated with the high spatial frequencies. The image 
    has a resolution of~0\farcs32. The contour levels are at
    1\%, 3\%, 5\%, 10\%, 30\% and 50\% of the brightness peak. The 
    background noise is $\sim 0.1\%$.}
  \label{HGL961SpeckFig}
\end{figure}

Figure~\ref{HGL961SpeckFig} shows the speckle reconstructed image of 
the double system formed by \glae\ and \glaw. In this case, the field of 
view is \fovs{14\farcs6}{14\farcs6}\ since the full
\fovp{256}{256}~pixel area of IRCAM3 was used. Both sources can be clearly
identified at a separation of \anypm{5\farcs29}{0\farcs03} and a 
PA\,=\,\angpm{17.9}{0.3}. A very faint third source 
(\glawb) can also be seen at a separation of \anypm{1\farcs48}{0\farcs06}\ and
a PA\,=\,\angpm{59}{3}\ from \glaw\ (renamed \glawa\ here). \glawa\ 
contributes a \anypm{14.2\%}{0.9\%} to the total flux, while the
contribution of \glawb\ is \anypm{1.0\%}{0.2\%}. We use the magnitudes given in
\citet{Castelaz85} to estimate a total magnitude in the H band of 8.65$\pm$0.01
for the system GL\,961-E, -Wa and -Wb. This yields a H magnitude of
10.8$\pm$0.1 for GL\,961-Wa and 13.6$\pm$0.1 for GL\,961-Wb. 

Some extended emission at the 1\% level can be seen 
around \glawa. The shape of the extended emission is fairly uncertain since 
it is very close to the noise level. In summary, there appears not to
be any clear feature in this source that can be related to any of the
outflow activity indicators seen at a larger scale.

High resolution observations of \fan\ were also performed at~K, since 
it may constitute an independent site of star formation. However, no
sub-arcsecond extended emission was found at a level $\la 1\%$\ of the
brightness peak.

\subsection[S235\,IRS1, 2 and 4]{S235\,IRS1, 2 and 4}

\begin{figure*}
  \centering
  \resizebox{\hsize}{!}{\includegraphics{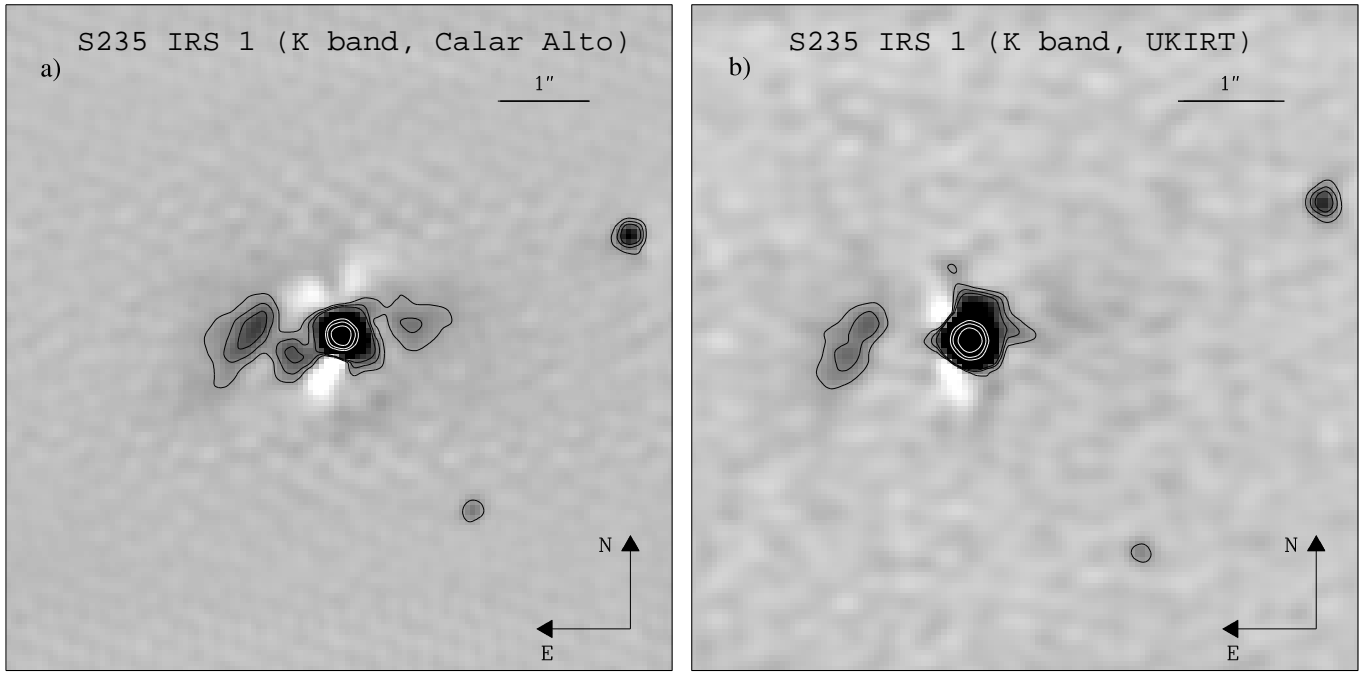}}
  \caption[Reconstructed image of \sone]{{\bf a)} Reconstructed image
    of \sone\ in the K~band taken with MAGIC
    during the Calar Alto campaign. The
    resolution achieved is 0\farcs27. In addition to IRS1,
    which is near the centre of the FOV, two other stars can
    be seen towards the NW and SW. The contours represent 
    0.5\%, 1\%, 1.5\%, 2\%, 3\% and 5\% of the peak. The noise level is
    $\sim 0.1\%$\ although negative artifacts at a level of 
    -2.5\% (white in the grey-scale) appear close to the
    brightness (central) peak. {\bf b)} The same as on panel a) but
    using the data taken during the campaign at UKIRT. In this case, 
    the negative artifacts are at a -1\% level.}
  \label{KS235IRS1BOTHFig}
\end{figure*}

These luminous YSOs \citep{Allen72} are located in the S235 molecular cloud
at a distance of 1.8~kpc. \sone\
was observed in speckle mode during 1994 at the Calar Alto observatory
in the K~band, and in 1997 at UKIRT in the H~and K~band. The K~band
speckle reconstructed images from these two observations at a resolution
of 0\farcs27 are shown in figures~\ref{KS235IRS1BOTHFig}a and
\ref{KS235IRS1BOTHFig}b.  In both K images, \sone\ can be seen near
the centre, and two other stars are detected to the west of IRS1. One
of them is located \anypm{4\farcs13}{0\farcs07}\ to the NW of \sone, and the
other \anypm{3\farcs0}{0\farcs1}\ to the SW. The brighter of these stars
contributes a \anypm{2.2\%}{0.1\%}\ to the total flux, while the
contribution of the fainter is \anypm{0.6\%}{0.1\%}. Assuming a total
magnitude of 10.25$\pm$0.07 ($11''$\ aperture,
c.f. \citealt{EvansBlair81}), the magnitudes of the bright and faint
stars are 12.29$\pm$0.08 and 13.7$\pm$0.2 respectively.

Some extended emission elongated in the E-W direction is clearly seen
in \sone.  The extended emission that is detected at
\anypm{1\farcs3}{0\farcs1}\ towards the east of the central peak, is
certainly real. The contribution to the total flux of this feature,
measured on the UKIRT image, is \anypm{6.2\%}{0.4\%}. However, the
features closer in are more questionable since there are negative
artifacts at a similar level.  Intrinsic time variability of the
nebulosity cannot be discarded, since the extended emission detected
to the west of IRS1 at Calar Alto (see figure~\ref{KS235IRS1BOTHFig}a) 
seems to have vanished in the 1997 data. 

The reconstruction of the H~band data yielded a low SNR image
due to the low level of counts on the reference star. Nevertheless,
inspection of the seeing-limited image indicates that the
nebulosity to the east of IRS1 is relatively bluer than the
brightness peak, which indicates the scattered nature of the
extension. Hence, we appear to have a monopolar or possibly bipolar
reflection nebula emanating from \sone\ in an E-W direction with the
blueshifted lobe on the east. Unfortunately, there are as yet not
molecular line maps with which to look for the outflow.  Near-IR
standard imaging in molecular hydrogen by M. McCaughrean (private
communication) shows some elongated diffuse emission in the E-W
direction extended for $\sim 1'$, which supports the picture of matter
outflowing in this direction.

\sfour\ it is associated with the optical nebulosity S235B
\citep{Krassner82}. The source lies $\sim 1'$\ south of the classical
\hii\ region S235A.  Centred on the region, there is also a large scale
bipolar outflow oriented in the NE-SW direction with the approaching
gas towards the NE \citep{BallyLada83,Nakano86}. \sfour\ was observed
in the H~band at UKIRT. The data show no evidence of sub-arcsecond
extended emission at a level of a 5\% of the brightness peak. The
source appears also unresolved in the K-band image from Calar Alto.
Little is also known of \sttft, which appears unresolved in both
images (H and K band) from the campaign at Calar Alto at the 5\% level.

\subsection[GL\,2591]{\glt}

This massive YSO is associated with an optical and near-IR nebula that 
extends for approximately $20''$\ to the west
\citep{Lenzen87,Yamashita87,Rolph91,Tamura91,Minchin91}. The source 
illuminating the nebula (IRS1), which is inferred from the 
centro-symmetry in the optical and near-IR polarisation pattern, is  
completely obscured in the optical.
The source is located near the centre of a large scale bipolar outflow
 oriented in the E-W direction with the blue-shifted lobe pointing to the
west \citep{Lada84,Mitchell92}. \citet{Yamashita87} detect extended CS 
emission in the N-S direction that can be interpreted as a large scale torus 
of dense material perpendicular to the outflow.

\begin{figure}
  \centering
  \resizebox{\hsize}{!}{\includegraphics{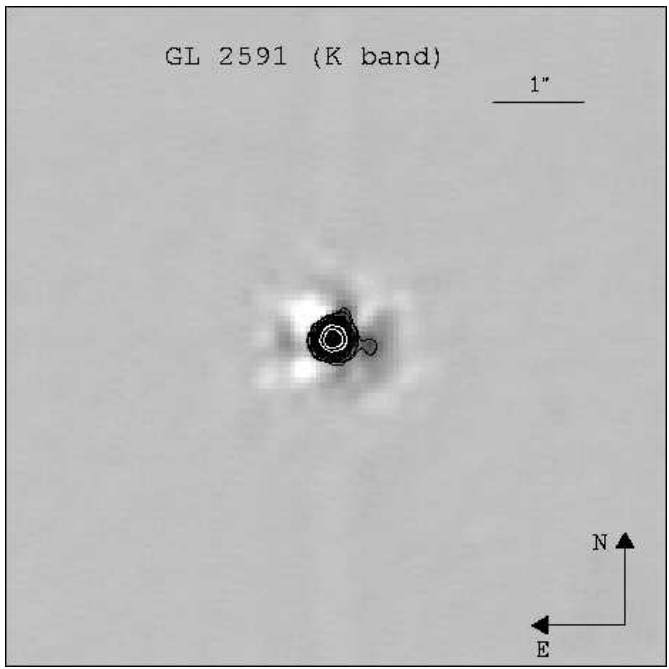}}
  \caption[Reconstructed image of \glt~IRS1 in the H~band]{Speckle 
    reconstructed image of \glt~IRS1 
    in the K~band at a resolution of 0\farcs21. There appears
    to be a faint extension to the west, at a distance of
    \anypm{0\farcs6}{0\farcs1} from the brightness peak. The
    contour levels are at 1.5\%, 2\%,
    5\%, 30\% and 50\% of the peak.  The noise level is $<1\%$\
    but the negative artifacts (white in the grey-scale) seen
    to the E of the brightness peak are at a -1.7\% level.}
  \label{GL2591SpeckleFig}
\end{figure}

\glt\ is associated with a faint radio source elongated in the E-W
direction \citep{GL2591Campbell84}. \citet{TamuraYamashita92} also
detect a group of \molh\ knots aligned in the E-W direction. Most of
the emission occurs towards the approaching lobe of the outflow,
although there is also some shock-excited gas to the west of IRS1. The
line of \molh\ knots to the~E of IRS1 appears to be continued by a
group of HH~objects aligned roughly along the same direction
\citep{Poetzel92}.


The speckle data of \glt\ taken at UKIRT show the presence of a faint knot
\anypm{0\farcs6}{0\farcs1} away at PA\,=\,\angpm{259}{8} with respect to the
brightness peak (figure~\ref{GL2591SpeckleFig}). The contribution to
the flux of is \anypm{1.3\%}{0.5\%}. This feature could be the base of
the larger scale reflection nebulosity to the west associated with the
blueshifted outflow lobe.

\subsection[GL\,4029\,IRS1]{\glftn}

   \begin{figure}
   \centering
   \resizebox{\hsize}{!}{\includegraphics{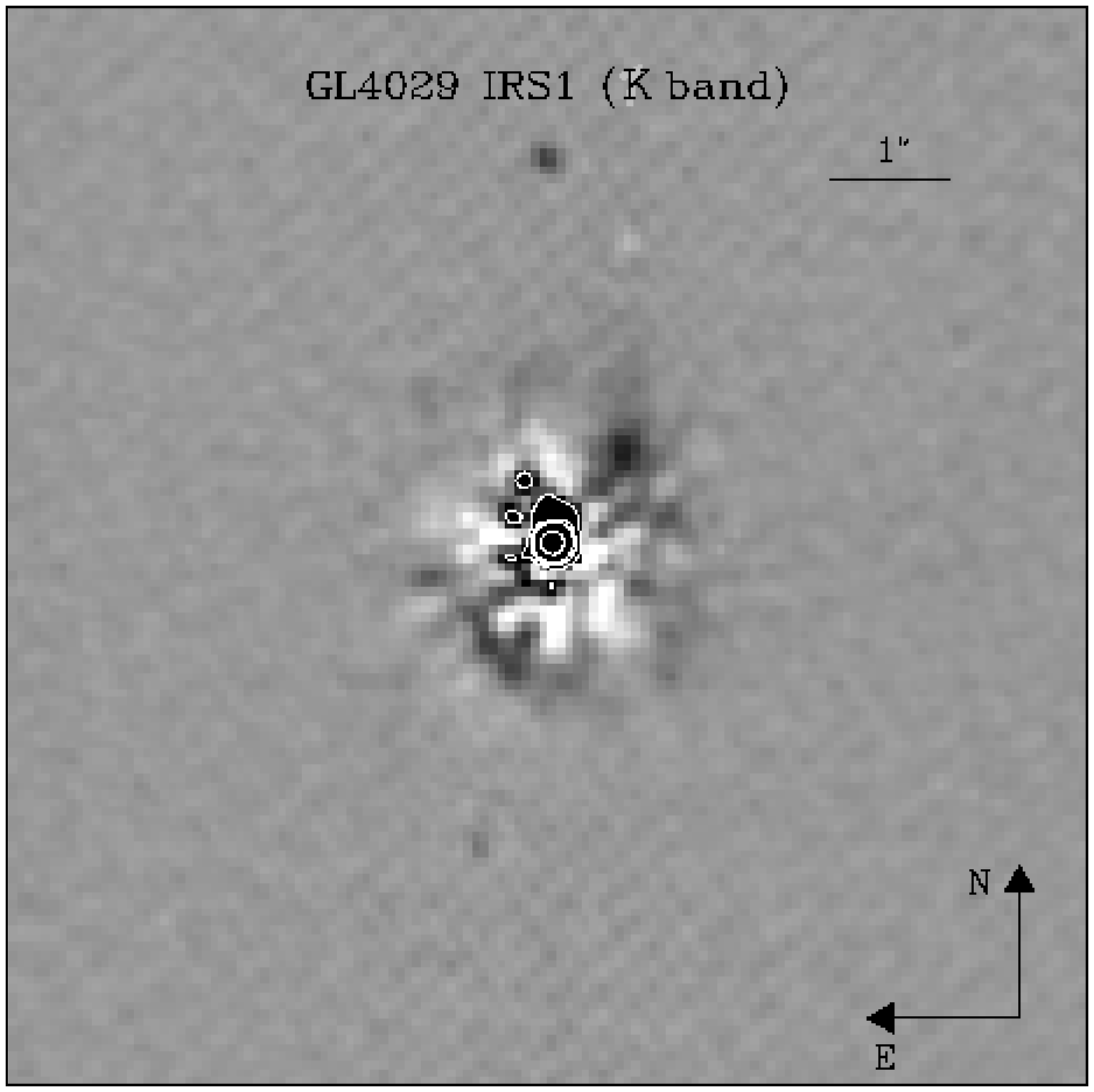}}
   \caption[Reconstructed image of \glftn\ in the K~band]{Speckle 
            reconstructed image of \glftn\ in the K~band at a
            resolution of 0\farcs19. The contour levels are at 1\%, 5\%,
            10\% and 50\% of the peak.  The noise level is $\la 1\%$. The 
            grey-scale varies from -0.8\% (white) to 2\% (black).}
    \label{GL4029SpeckleFig}
    \end{figure}

This massive YSO is a member of a cluster of IR sources
\citep{Deharveng97} located at a distance of 2.2~kpc
\citep{BeckerFenkart71}. \glftn\ is located near the centre of a
bipolar molecular outflow \citep{Snell88}. Both outflow lobes have
their peak emission coincident with the IR source. This maybe caused
either by a low degree of collimation or by the outflow pointing
towards us.  \glftn\ has an associated reflection nebula in
the optical and near-IR \citep{Deharveng97}.  \citet{RayPoetzel90}
found several
optical emission line knots emanating from \glftn\ that indicate the
presence of a jet at a \pa{255} with a projected extension of
$13''$. High resolution observations with the VLA \citep{Zapata01}
resolve the emission at 3.6~cm into two components separated by
0\farcs6 in the N-S direction with a possible counterpart to the
optical jet being detected in the southern component. Sub-arcsecond
resolution mid-IR observations also show some extended emission
emerging from \glftn\ \citep{Zavagno99}, which appears to be associated
with the jet.

Recent HKL$^{\prime}$ 1D speckle interferometry by \citet{Leinert01}
show the presence of an extended scattered light halo of diameter
\anypm{1\farcs4}{0\farcs3}. The K-band speckle image taken at Calar Alto
(figure~\ref{GL4029SpeckleFig}) shows some signs of such an extended
halo, although it is somewhat marred
by artifacts from the reconstruction procedure. Similarly, our K-band
visibility function could be seen as being consistent with a 20\%
extended halo contribution, but unfortunately the visibility function
of one reference star with respect to the other shows a similar drop
at high spatial frequencies and so is not conclusive. However, it is
interesting to note that \citet{Leinert01} deduced the presence of a
patch of separate diffuse emission to the NW of the central source from their
1D data and this is exactly where we see the most convincing part of
the halo in our K-band image. We also see hints of some extended
emission just to the north of \glftn. Neither of these correspond 
with the jet direction and so their nature is unclear. 

Besides IRS1 there are two more stars, one towards the north at a
distance of \anypm{3\farcs1}{0\farcs1}\ and a PA of \angpm{1.27}{0.03}\ and
the other towards the south at a distance of \anypm{2\farcs6}{0\farcs1}\ and
a PA of \anypm{166.04}{0.04}. Two emission features are seen towards
the NE within $1''$\ of IRS1, but their nature is questionable and neither
exactly coincides within the positional errors with the northern radio 
source found by \citet{Zapata01}.

\subsection[NGC\,7538\,IRS1]{\ngcsfte}

\ngcsfte\ is the brightest of a cluster of IR sources located in the 
southern edge of the optical \hii\ region NGC\,7538 
\citep{Wynn-WilliamsBecklin74, Tamura91}. The cluster is at 
the centre of a bipolar molecular outflow at a \pa{-45}\ that extends
for \anyapp{1}{pc}\ with the blue-shifted lobe towards the NW
\citep{Fischer85,Scoville86,Kameya89,DavisHoare98}.

VLA observations at 5 and \ghz{15}\ of \ngcsfte\
\citep{NGC7538Campbell84} show that the compact ionised gas emission
has a bipolar morphology elongated in the N-S direction (\pa{165 -
180}). They interpret this emission as the result of a collimated
wind. \citet{Gaume95} found that the $\mathrm{H66\alpha}$\
recombination lines were very broad, indicating a wind speed of
$\sim$~\kms{250}. \citet{DavisHoare98}\ found shock excited \molhl\
emission extending about an arcminute northwards of \ngcsfte
\citep[see also][]{Bloomer98}.  Bow-shaped features are also seen in
the vicinity of IRS1, but it is difficult to uniquely identify their 
powering source.

\begin{figure}
  \centering
  \resizebox{\hsize}{!}{\includegraphics{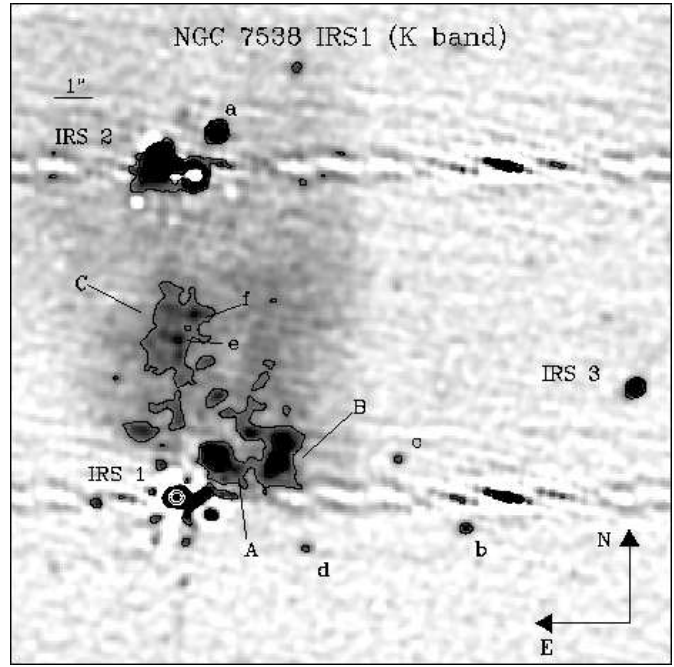}}
  \caption[Reconstructed image of \ngcsfte\ in the K~band]{Speckle 
    reconstructed image of \ngcsfte\ in the K~band at a
    resolution of $0\farcs33$. The contour levels are at 0.3\%, 0.5\%,
    1\%, 2\%, 25\% and 50\% of the peak.  The noise level is
    $\sim 0.2\%$. The grey-scale varies from -0.1\% (white) to 
    1\% (black). In this case, the full FOV of MAGIC was used, and
    therefore the field of view is \fovs{17\farcs9}{17\farcs9}.
    The two parallel horizontal stripes that cross the image 
    at the positions of IRS1 and IRS2 are artifacts from the 
    reconstruction.}
  \label{NGC7538SpeckleFig}
\end{figure}

\citet{Scoville86} found a dense core of \thco\ emission towards
\ngcsfte\ that is elongated in the E-W direction. The 
detection of a line of methanol masers that extends for
\anyapp{300}{AU}\ at a \pa{110}\ (roughly perpendicular to the outflow 
direction) may indicate the presence of an edge-on disc in this source 
\citep{Minier01}.

The region shows also a large scale near-IR nebulosity
(\anyapp{0.3}{pc}) which has a monopolar shape extending towards the 
blue-shifted lobe of the large scale molecular outflow with \ngcsfte\
at its apex \citep{Tamura91}. Its centro-symmetric polarisation pattern 
indicates that \ngcsfte\ is the illuminating source.

Our K~band reconstructed image of \ngcsfte\ has a complex structure (see
figure~\ref{NGC7538SpeckleFig}).  Due to the presence of several
bright sources in the field the artifacts are particularly strong
here, especially the two parallel horizontal stripes that cross the
image at the positions of IRS1 and IRS2. In addition to IRS1, IRS2 and IRS3,
other near-IR sources also appear in the \fovs{17\farcs9}{17\farcs9}\ field
of view of the image. The stars labelled from 'a' to 'd' also appear in
the seeing-limited image of \citet{Bloomer98}. The stars 'e' and 'f'
however, are not seen in their image. The extended features 'A' to
'C', which are part of the large scale near-IR reflection nebula,
appear as regions of enhanced emission in the image presented by
\citet{Bloomer98}. 

It is notable that the extended emission is mainly concentrated
towards the N and NE of IRS1, which is roughly the orientation of the
blue-shifted lobe in the CO outflow. Features 'B' and 'C' would appear
to be the result of scattered light in the walls a limb-brightened
cavity at a \pa{-35}\ with its apex at IRS1 and with an opening angle
of $\sim $\degg{70}. However, no bright sub-arcsecond nebulosity is
seen that would correspond to the base of this large scale cavity.

\subsection[GL\,5180\,IRS1]{\gfive}

\begin{figure}
  \centering
  \resizebox{\hsize}{!}{\includegraphics{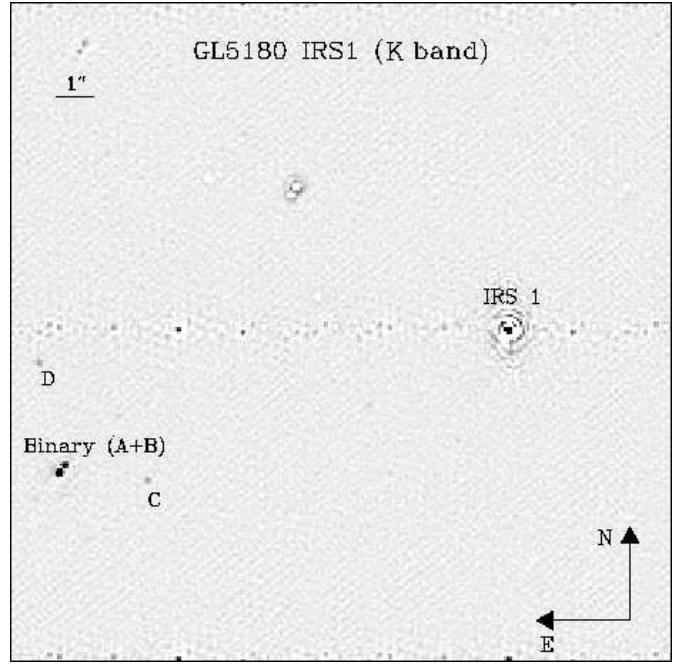}}
  \caption[Reconstructed image of \gfive\ in the K~band]{Speckle 
    reconstructed image of \gfive\ in the K~band at a
    resolution of 0\farcs07. The noise level is $\sim 0.5\%$
    although near IRS1 there are noise features with a value 
    of $-15\%$. The grey-scale varies from -1\% (white) to 
    8\% (black). In this case the full array was used, and
    therefore the field of view is \fovs{18''}{18''}. The image
    shows several artifacts produced during the reconstruction. These
    appear on the row through IRS1 and as positive and negative
    artifacts of the binary images.}
  \label{GL5180peckleFig}
\end{figure}

\gfive\ is near the centre of a CO outflow oriented at a \pa{130}\
\citep{Snell88}.  K-band imaging polarimetry shows a small cluster
within extensive nebulosity \citep{Tamura91, Yao00}. 
figure~\ref{GL5180peckleFig} shows our reconstructed image of
\gfive\ in the K~band covering a field of view of \fovs{18''}{18''}. 
We find a new binary at a distance of
\anypm{12\farcs8}{0\farcs1}\ and \angpm{107.5}{0.1}\ with respect to IRS1. 
The two binary components (A and B in figure~\ref{GL5180peckleFig}) are
separated by \anypm{0\farcs3}{0\farcs1}. The component B is at a PA of 
\angpm{320}{1}\ with respect the component A. Two other 
IR point sources are seen at \anypm{10\farcs6}{0\farcs1}\ and PA of
\angpm{113}{1}\ (source C) and at \anypm{12\farcs8}{0\farcs1}\ and PA of 
\angpm{94}{1}\ (source D). None of these sources (except for
IRS1) coincides with the IR sources found by \citet{Tamura91}. No 
extended emission is detected.  

\subsection[GL\,989]{\glb}
 
\glb\ (NGC~2264~IRS) is an infrared source discovered by
\citet{Allen72}, which is located at a distance of 750 pc.  
The presence of a molecular outflow in the region was
first described by \citet{BallyLada83}. The molecular outflow
structure was studied in more detail by \citet{Schreyer97}. Their CO
map appears to show that the outflow is oriented rather pole-on. The
IR source lies at the apex of a fan-shaped nebula with an approximate
visual magnitude of 15, extending to the NW \citep{Schmidt72}. The
morphology of the nebula was studied in the optical by
\citet{Scarrott89}. They observed a colour gradient characteristic of
reflection nebulae, i.e. it is redder the closer to the compact IR
source. However, the absence of centro-symmetry around \glb\ in the
polarisation pattern, indicates that the nebulosity is not simply a
reflection nebula illuminated by this source.  \citet{Schreyer97}
studied the morphology of the extended nebula in the near-IR. In this
wavelength range, the nebula is also more extended towards shorter
wavelengths, which is consistent with a scattered origin for the
emission.

Our H-band visibility function obtained at Calar Alto indicated that
some 20\% of the flux was extended on scales larger than
1\arcsec. Hints of diffuse emission to the NW are seen in the
reconstructed image as expected if we are detecting the larger scale
nebula, but no sub-arcsecond structure was seen. Repeated observations
with UKIRT at both H and K-bands failed to detect any extended
emission at the 4\% level.  High resolution observations with NICMOS
on board the Hubble Space Telescope \citep{Thompson97} also failed to
detected any extended emission.

\subsection[NGC\,2024\,IRS2]{\ngctwo}

No nebulosity was found around this object, but a fairly bright
companion was found 5\farcs1 to the east (PA\,=\,283$^\circ$). It has a
flux at K of about 10\% that of NGC\,2024\,IRS2 and 20\% at H. Hence, it
is somewhat bluer than IRS2, but still a very red object. This object
is not mentioned in the mid-IR survey of \citet{Haisch01}. 

\subsection[Mon\,R2\,IRS2]{\mont}

This source is embedded in the \monr\ \hii\ region itself and the strong
background nebulosity caused some problems in the analysis. However,
the source does not appear to have any sub-arcsecond
structure. Inspection of the radially averaged visibility function
yields an upper limit on any contribution to the total flux of about
10\% level.

\subsection[S106\,IR]{\sir}

No sub-arcsecond structure was found in this object with upper limits
on a contribution to the total flux of about 3\% at K and 6\% at
H. Possible reflected features are seen extending with a wide opening
angle a few arcseconds to the south and west in the $r$-, $i$- and
$z$-bands by \citet{Persson88}, but their nature
has not been confirmed polarimetrically.  Recent near-IR Adaptive Optics
imaging of this source shows no sub-arcsecond structure
\citep{Feldt02}. \citet{Puga03} present a K$'$\ polarisation map 
  of this region at a resolution of 0\farcs3. No conclusive proof of
  the presence of the extended features found by Persson et al. can be
  seen in this map. No clear polarisation pattern is observed within 
  a few arcseconds of \sir.

\section[Discussion]{Discussion}
\label{DiscussionSection}

Six (\glfts, \mon, \glf, S255 IRS3, \stwo, S235 IRS1) out of the 21
massive YSOs observed show a
significant small scale monopolar reflection nebula above the
$\sim$1\% dynamic range of the speckle imaging presented here. In
these cases, the nebular morphology is consistent with light being
scattered off the dust in the walls of an outflow cavity. There is
usually good agreement between the position angle of the sub-arcsecond
nebula and the larger scale outflow indicators such as CO and larger
near-IR nebulae. This agreement should be taken with
caution though, since in some instances, large scale CO flows appear 
as multiple outflows in interferometric studies \citep[e.g. ][]{Beuther03}.

We now address the question of why only 28\% of the sources show
nebulosity at this level and the others do not, bearing in mind that
the sample was not selected in any systematic way.  Simulations show
that by far the most important parameter for the brightness of a
reflection nebula for YSOs is the overall optical depth of the
envelope \citep[e.g. ][]{WhitneyHartmann92,Fischer94,Lucas98,Alvarez04}.  
Too high an optical depth and the source is either totally invisible or
only a faint nebula is present with no bright point source
with which to utilise the speckle imaging technique. When the optical depth
becomes very low, the central star will dominate the near-IR light such
that any nebulosity will be lost in the limited dynamic range of
speckle imaging. Other factors such as the inclination angle, the
opening angle of the cavity, the density distribution in the envelope
and the dust parameters also play a role in determining the nebular
brightness. However, these are secondary
  if the density scale is altered such that the line of sight optical
  depth is kept constant \citep{Alvarez04}.

The picture outlined above would primarily be an evolutionary one
with the circumstellar matter being gradually
cleared by the outflow, and hence there will be less dust available
for scattering. There will also be less dust obscuring the central
star, and therefore the contrast between the reflection nebula and the
central source will increase. At some point the reflection nebulae
become undetectable at the 1\% level. In those objects where there is 
insufficient optical depth at K to detect the nebula, then going to
shorter wavelengths should reveal the nebulae (e.g. \glf), as the 
scattering efficiency increases and the star is obscured. Unfortunately, 
this can often not be possible with speckle imaging, as the nebula becomes 
also too faint, even though it is relatively brighter than the star.

\begin{figure}
  \centering
  \resizebox{\hsize}{!}{\includegraphics{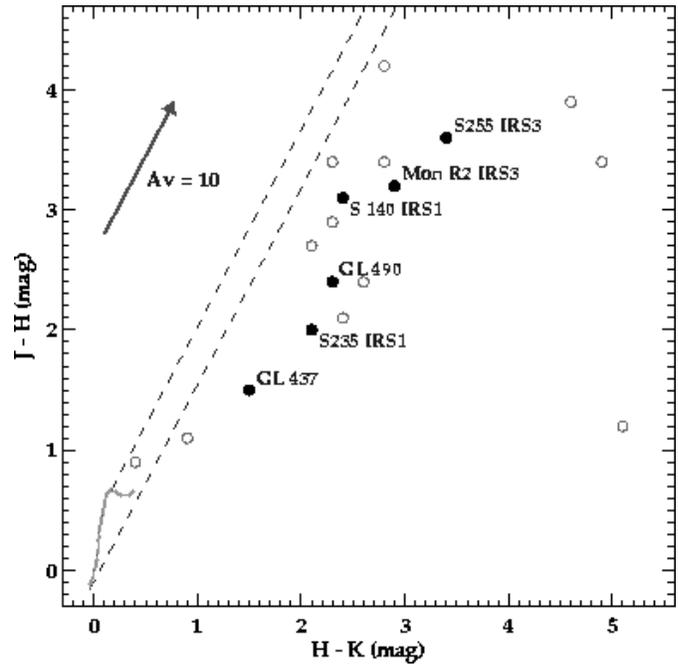}}
  \caption[Colour-Colour diagram]{Colour-colour diagram for the
    massive YSOs observed in our speckle campaigns. The filled
    circles represent the sources in which a sub-arcsecond conical
    nebula was detected. The open circles represent sources in which 
    no clear extended emission associated with the outflow was
    detected. Note that our typical detection limit is at 1\% of the 
    brightness peak. The main sequence is plotted as a grey line. The 
    dashed lines indicate the extincted main sequence using the
    extinction law of \citet{He95}. The arrow represents a visual
    extinction of 10 magnitudes.}
  \label{CavityODFig}
\end{figure}

To investigate the detection dependence on optical depth we have plotted our
targets on a $J-H$\ versus $H-K$\ colour-colour diagram in
figure~\ref{CavityODFig}. There
is no clear separation of those with bright sub-arcsecond nebulae and
those without in this diagram. The most we can say is that the very
lightly reddened sources are not showing the phenomena.  This is
consistent with these simply not being embedded enough to still
produce bright nebulae relative to their central star. However, not
all the detections are among the most reddened sources. Of course,
using such a colour-colour diagram as an indicator of the optical
depth of the circumstellar envelope is fraught with many
complications. Some of the highly reddened non-detections could simply
suffer from high line-of-sight extinction that does not arise in the
immediate circumstellar envelope of the object itself. The K-band
brightness is likely to be dominated by hot circumstellar dust
contributions in many cases whilst the shorter wavelengths can be
primarily due to unresolved scattered light.

We might expect a trend with distance in that it is easier to detect 
the cavity base in the nearby objects. However, there is not
any indication of this in the limited sample we have studied here.

Although the aim of this study was not to search for close companions
in massive YSOs, it is worth to note that 3 out of the 21 sources show
close companions (\lkhao, \mon, \gla -Wa) with separations between
0\farcs17 and 1\farcs3 . These angular separations correspond to
projected distances between 150~AU and 1800~AU. This represents a
binary detection frequency of 14$\pm$8\%. The statistical
significance in our sample is fairly poor though, and more massive
YSOs should be studied before a robust estimate on the frequency of
binaries in massive YSOs can be achieved. In any case, it is worth
comparing our binary detection rate with that found in Herbig Ae/Be
and OB stars. In a sample of 31 Herbig Ae/Be stars, \citet{Leinert97}
find that $\sim$\,16\% have sub-arcsecond separations and $\sim$\,29\%
have projected separations $<$\,1000~AU. \citet{Duchene01} find, in a
sample of 60 OB stars, an 18\%$\pm$6\% sub-arcsecond binary detection 
rate (projected separations in the range 200~AU -- 3000~AU). Hence,
we find a binary detection frequency that is in good agreement with that
found in Herbig Ae/Be and OB stars. 
  
\section[Conclusions]{Conclusions}
\label{ConclusionsSection}

The sub-arcsecond morphology in a sample of 21 massive YSOs has been
studied using speckle imaging in the near-IR.  About 30\% of the
sources show a sub-arcsecond monopolar reflection nebula.  These
nebulae are consistent with being due to light scattered off the walls
of the cavity evacuated by the blueshifted lobe of a bipolar outflow
in most cases. A wide range of opening angles is seen and some have a
gap between the star and the base of the nebular cone. The size of the
sample is still far from that necessary to examine trends in opening
angle such as whether the cavity widen in angle in older sources as
the outflow clears the envelope. Larger and well-selected samples of
massive YSOs will be needed to address the properties and evolution of
the outflow cavities systematically \citep[see ][]{Lumsden02}.

For the majority of sources where no sub-arcsecond nebulosity is seen,
it could be due to them being at a slightly more evolved stage where
their circumstellar envelope is not sufficiently optically thick. This
could be checked by resolving the emission from the warm or cool dust
in these envelopes at either mid-IR or sub-millimetre wavelengths. A
In a related paper \citep{Alvarez04}, we use a Monte Carlo code of the 
scattered light to examine in detail the detectability and the
dependence of the sub-arcsecond nebulae on the physical properties of 
the circumstellar matter.

Adaptive optics observations are capable of much
higher dynamic ranges and will therefore be able to detect a much
higher fraction of nebulae than was possible here with speckle
imaging. This will also help test the scenario in which the reflection
nebulae fade as the outflow clears material from the envelope. Larger
telescopes will allow the cavity to be traced closer to the star 
where collimation of the flow is likely to be occurring.

\begin{acknowledgements}
      	CA would like to thank to the Physics and Astronomy Department
      	at Leeds University for their support. CA is also deeply
      	grateful to Kapteyn Astronomical Institute for allowing him to 
	use their facilities during the realization of this work. We
      	thank the anonymous referee for the useful comments
      	and suggestions. This research has made use of NASA's
      	Astrophysics Data System.   
\end{acknowledgements}


\end{document}